%% file: Lillo-Box_K2-32_K2-233_PlanetarySystems.tex
\documentclass{aa}  
\usepackage{graphicx}
\usepackage{txfonts}
\usepackage{graphicx}
\usepackage{lscape}
\usepackage{longtable}
\usepackage{natbib}
\usepackage{color}
\usepackage{array} 
\usepackage{array} 
\usepackage{tikz,array}
\usetikzlibrary{calc}
\usepackage{txfonts}
\usepackage{color}
\usepackage{multirow}
\usepackage{here}

\usepackage{txfonts}
\usepackage{amsmath,amstext}
\usepackage{graphicx}
\usepackage{epstopdf}
\usepackage{float}
\usepackage{array}
\usepackage{mathtools}
\usepackage{booktabs}
\usepackage{subfigure}
\usepackage{url}
\usepackage{helvet}
\usepackage{tabularx}
\usepackage{multirow}
\usepackage{natbib}
\usepackage[flushleft]{threeparttable}
\usepackage{lscape}
\usepackage{pdflscape}
\usepackage{longtable}
\usepackage{wasysym}
\usepackage{float}
\usepackage{relsize}
\usepackage{color}
\usepackage{breqn}
\usepackage{bm}

\usepackage[colorlinks, citecolor=blue, linkcolor=blue, breaklinks=true]{hyperref}
\hypersetup{colorlinks,breaklinks, linkcolor=blue,urlcolor=magenta, anchorcolor=blue,citecolor=blue}

\usepackage{scalerel}
\usepackage{tikz}
\usetikzlibrary{svg.path}
\definecolor{orcidlogocol}{HTML}{A6CE39}
\tikzset{
  orcidlogo/.pic={
    \fill[orcidlogocol] svg{M256,128c0,70.7-57.3,128-128,128C57.3,256,0,198.7,0,128C0,57.3,57.3,0,128,0C198.7,0,256,57.3,256,128z};
    \fill[white] svg{M86.3,186.2H70.9V79.1h15.4v48.4V186.2z}
                 svg{M108.9,79.1h41.6c39.6,0,57,28.3,57,53.6c0,27.5-21.5,53.6-56.8,53.6h-41.8V79.1z M124.3,172.4h24.5c34.9,0,42.9-26.5,42.9-39.7c0-21.5-13.7-39.7-43.7-39.7h-23.7V172.4z}
                 svg{M88.7,56.8c0,5.5-4.5,10.1-10.1,10.1c-5.6,0-10.1-4.6-10.1-10.1c0-5.6,4.5-10.1,10.1-10.1C84.2,46.7,88.7,51.3,88.7,56.8z};
  }
}
\newcommand\orcidicon[1]{\href{https://orcid.org/#1}{\mbox{\scalerel*{
\begin{tikzpicture}[yscale=-1,transform shape]
\pic{orcidlogo};
\end{tikzpicture}
}{|}}}}
\def\kms{km s$^{-1}$}         %km.s -1
\def\ms{\hbox{m s$^{-1}$}}         %m.s -1
\def\gcm3{\hbox{g cm$^{-3}$}}       %g.cm-3
\def\Msun{\hbox{$\mathrm{M}_{\astrosun}$}}             %Msun
\def\Rsun{\hbox{$\mathrm{R}_{\astrosun}$}}

\def\Mearth{\hbox{$\mathrm{M}_{\oplus}$}}
\def\Rearth{\hbox{$\mathrm{R}_{\oplus}$}}
\def\degr{\hbox{$^\circ$}}
\def\teff{T$_{\rm eff}$}
\def\logg{log~{\it g}}
\def\met{[Fe/H]}

\bibpunct{(}{)}{;}{a}{}{,} % to follow the A&A style

\newcommand{\be}{\begin{equation}}
\newcommand{\ee}{\end{equation}}

\newcommand\mearth{{M$_{\oplus}$}}
\newcommand\rearth{{R$_{\oplus}$}}
\newcommand\pastis{{\texttt{PASTIS}}}

\usepackage{romannum}

\usepackage{graphicx}
\usepackage{txfonts}
\begin{document}

   \title{Masses for the seven planets in K2-32 and K2-233\thanks{Based on observations collected at the European Organisation for Astronomical Research in the Southern Hemisphere under ESO programmes 198.C-0169 and 095.C-0718.}\, \thanks{The full version of Tables \ref{tab:k2-32_lc}, \ref{tab:k2-233_lc}, \ref{tab:k2-32_rv}, and \ref{tab:k2-233_rv} are available in electronic form at the CDS via anonymous ftp to cdsarc.u-strasbg.fr (130.79.128.5) or via \url{http://cdsweb.u-strasbg.fr/cgi-bin/qcat?J/A+A/}}}

   \subtitle{Four diverse planets in resonant chain and the first young rocky worlds}

   \author{
%%--------------Lead Authors-----------------------
   J.~Lillo-Box\inst{\ref{cab}}, %\orcidicon{0000-0003-3742-1987}, 
   T.~A.~Lopez\inst{\ref{marseille}}, %\orcidicon{0000-0001-6622-1250},
   A.~Santerne \inst{\ref{marseille}}, %\orcidicon{0000-0002-3586-1316},
%------------Contributing authors -----------------
   L.~D.~Nielsen\inst{\ref{geneva}}, %\orcidicon{0000-0002-5254-2499},
   S.C.C.~Barros\inst{\ref{porto}}, %\orcidicon{0000-0003-2434-3625},
   M.~Deleuil\inst{\ref{marseille}}, 
   L.~Acu\~na\inst{\ref{marseille}}, %\\
   O.~Mousis\inst{\ref{marseille}}, 
   S.~G.~Sousa\inst{\ref{porto}},%\orcidicon{0000-0001-9047-2965},
   V.~Adibekyan\inst{\ref{porto},\ref{porto2}}, %\orcidicon{0000-0002-0601-6199},
   D.~J.~Armstrong\inst{\ref{ceh}}\fnmsep\inst{\ref{warwick}}, %\orcidicon{0000-0002-5080-4117},
%------------Alphabetical order -----------------
   D.~Barrado\inst{\ref{cab}},%\orcidicon{0000-0002-5971-9242},
   D.~Bayliss\inst{\ref{ceh}}\fnmsep\inst{\ref{warwick}}, %\\%\orcidicon{0000-0001-6023-1335},\\
   D.~J.~A.~Brown\inst{\ref{ceh}}\fnmsep\inst{\ref{warwick}}, %\orcidicon{0000-0003-1098-2442},
   O.D.S.~Demangeon\inst{\ref{porto}}, %\orcidicon{0000-0001-7918-0355},
   X.~Dumusque\inst{\ref{geneva}}, %\orcidicon{0000-0002-9332-2011},
   P.~Figueira\inst{\ref{eso},\ref{porto}}, %\orcidicon{0000-0001-8504-283X},
   S.~Hojjatpanah\inst{\ref{porto},\ref{porto2}}, %\\%\orcidicon{0000-0002-0417-1902},\\
   H.~P.~Osborn\inst{\ref{marseille}}\fnmsep\inst{\ref{bern}}\fnmsep\inst{\ref{mit}}, %\orcidicon{0000-0002-4047-4724},
   N.~C.~Santos\inst{\ref{porto},\ref{porto2}}, %\orcidicon{0000-0003-4422-2919},
   S.~Udry\inst{\ref{geneva}}, %\orcidicon{0000-0001-7576-6236}
   }

\institute{
Centro de Astrobiolog\'ia (CAB, CSIC-INTA), Depto. de Astrof\'isica, ESAC campus 28692 Villanueva de la Ca\~nada (Madrid), Spain\label{cab} \email{Jorge.Lillo@cab.inta-csic.es  }
\and Aix Marseille Univ, CNRS, CNES, LAM, Marseille, France \label{marseille} 
\and Geneva Observatory, University of Geneva, Chemin des Mailettes 51, 1290 Versoix, Switzerland \label{geneva}
\and Instituto de Astrof\' isica e Ci\^encias do Espa\c{c}o, Universidade do Porto, CAUP, Rua das Estrelas, PT4150-762 Porto, Portugal \label{porto} 
\and Departamento de F\'isica e Astronomia, Faculdade de Ci\^encias, Universidade do Porto, Rua do Campo Alegre, 4169-007 Porto, Portugal \label{porto2}
\and Centre for Exoplanets and Habitability, University of Warwick, Gibbet Hill Road, Coventry, CV4 7AL, UK \label{ceh} 
\and  Department of Physics, University of Warwick, Gibbet Hill Road, Coventry CV4 7AL, UK \label{warwick}
\and  NCCR/Planet-S, Centre for Space and Habitability, University of Bern, Bern 3012, Switzerland \label{bern} 
\and  Kavli Institute for Astrophysics and Space Research, Massachusetts Institute of Technology, Cambridge, MA 02139, USA \label{mit} 
\and European Southern Observatory, Alonso de Cordova 3107, Vitacura, Region Metropolitana, Chile \label{eso}
}

\titlerunning{K2-32 and K2-233}
\authorrunning{Lillo-Box et al.}

   \date{in prep.}

% \abstract{}{}{}{}{} 
% 5 {} token are mandatory
 
  \abstract
   {High-precision planetary densities are key pieces of information necessary to derive robust atmospheric properties for extrasolar planets. Measuring precise masses is the most challenging part of this task, especially in multi-planetary systems. The ESO-K2 collaboration has followed-up a selection of multi-planetary systems detected by the K2 mission using the HARPS instrument with this goal in mind.}
% in preparation for JWST
   {In this work we measure the masses and densities of two multi-planetary systems: a four-planet near resonant chain system (K2-32), and a young  ($\sim400$ Myr old) planetary system  consisting of three close-in small planets (K2-233).}
   {We have obtained 199 new HARPS observations for K2-32 and 124 for K2-233 covering a long baseline of more than three years. We perform a  joint analysis of the radial velocities and K2 photometry with \texttt{PASTIS} to precisely measure and constrain the properties of these planets, focusing on their masses and orbital properties.}
   {We find that K2-32 is a compact scaled-down version of the Solar System's architecture, with a small rocky inner planet (M$_e=2.1^{+1.3}_{-1.1}$~\Mearth, P$_e\sim4.35$~days) followed by an inflated Neptune-mass planet (M$_b=15.0^{+1.8}_{-1.7}$~\Mearth, P$_b\sim8.99$~days) and two external sub-Neptunes (M$_c=8.1\pm2.4$~\Mearth, P$_c\sim20.66$~days; M$_d=6.7\pm2.5$~\Mearth, P$_d\sim31.72$~days). K2-32 becomes one of the few multi-planetary systems with four or more planets known where all have measured masses and radii. Additionally, we constrain the masses of the three planets in the K2-233 system through marginal detection of their induced radial velocity variations. For the two inner Earth-size planets we constrain their masses {at a 95\% confidence level} to be smaller than  M$_b<11.3$~\Mearth{} (P$_b\sim2.47$~days), M$_c<12.8$~\Mearth{} (P$_c\sim7.06$~days). The outer planet is a sub-Neptune size planet with an inferred mass of M$_d=8.3^{+5.2}_{-4.7}~\Mearth$ (M$_d<21.1$~\Mearth{}, P$_d\sim24.36$~days).   }
   {Our observations of these two planetary systems confirm for the first time the rocky nature of two planets orbiting a young star, with relatively short orbital periods ($<7$ days). They provide key information for planet formation and evolution models of telluric planets. Additionally, the Neptune-like derived masses of the three planets K2-32\,b,\,c,\,d puts them in a relatively unexplored regime of incident flux and planet mass, key for transmission spectroscopy studies in the near future.}


   \maketitle
%
%________________________________________________________________
\section{Introduction}

After the prime \textit{Kepler} quest \citep{borucki10}, the extension of the mission, \textit{K2} \citep{howell14}, explored 19 fields along the ecliptic, mainly focusing on a search for transiting planets. Each of these campaigns observed a single 100 deg$^2$  field over 80 days, monitoring the brightness of about 20,000 stars with a 30-min cadence. Overall a larger number of bright targets were observed compared to the primary mission, including several well-known young stellar associations {such as} the Pleiades, Praesepe {and} the Hyades. To date, more than 390 planets have been detected by \textit{K2} and more than 900 remain as candidates awaiting confirmation (see NASA Exoplanet Archive\footnote{\url{https://exoplanetarchive.ipac.caltech.edu/docs/counts_detail.html}}, \citealt{akeson13}).
Among the confirmed planets, \textit{K2} and associated ground-based follow-up have covered planets in very different regimes, from a system with one of the longest-known resonant chain configurations (K2-138, \citealt{lopez19}), super-Earths and mini-Neptunes in the habitable zone of their parent stars (e.g. K2-18\,b, \citealt{montet15}), planets in young stellar associations (K2-95\,b, \citealt{obermeier16,pepper17}) and even an Earth-size planet with Mercury-like composition (K2-229\,b, \citealt{santerne18}).

The detection of these transiting planets by \textit{K2} is only the first step towards a full characterisation that puts them in context with the current plethora of known exoplanets. {The combination of transit and high-precision radial velocity (RV) analysis provides the absolute planet mass and mean density}, a key parameter to understand planetary internal structure \citep[e.g., ][]{dorn15}. Transiting planets also give us the ability to study the atmospheric properties of planets through different techniques like transmission spectroscopy \citep{seager00}. However, \cite{batalha19} found that precisions better than 20\% in the planet mass are necessary to obtain accurate atmospheric properties, especially for low-mass (sub-Neptune) planets. Only long-term dedicated follow-up programs on high-precision spectrographs can achieve this critical goal. 

Our Large Programme (PI: A. Santerne) on the ESO/HARPS instrument \citep{mayor03} focused on obtaining precise mass measurements of a sample of 25 transiting planets in 9 planetary systems detected by the K2 mission, with the ultimate goal of bridging the gap between rocky and gaseous planets by measuring densities to a precision of 30\% or better. In total, 18 planets have already been published from this programme: HD106315 b \& c \citep{barros17}, K2-229 b, c, \& d \citep{santerne18}, K2-265 b \citep{lam18}, K2-138b, c, d, e, f, g \citep{lopez19}, and HIP\,41378 b, c, d, e, f \& g \citep{santerne19}.

In the case of multi-planetary systems, {access to planet bulk compositions also enables studies of their architecture and subsequently their history of formation and evolution}. In this sense, the pile-up of planets in or near resonant chains provides information about the migration history of the planetary system, a direct consequence of convergent migration \citep[see][and references therein]{delisle17}. Also relevant is that these multi-planetary systems in resonant chains are the preferred environments to look for co-orbital planets \citep[e.g.,][]{lillo-box18a, leleu19}, since these resonances increase the stability of 1:1 co-orbital configurations \citep{leleu19b}, commonly formed in simulations of the formation and early evolution of multi-planet systems \citep{cresswell08, leleu19b}. Multi-planetary systems are also unique laboratories to test the diversity of planet properties among planets grown in the same environment. 

Four planets were {discovered} by the \textit{K2} mission in the \object{K2-32} system \citep{sinukoff16,heller19}. The planets were statistically validated and their masses were estimated for planet $b$ and upper limits found for the masses of the planets $c$ and $d$ using data from different high-resolution spectrographs \citep{dai16,petigura17}. No mass {constraint} has been published to date on the inner planet K2-32\,e. We have now collected 199 new HARPS measurements that we analyze in this paper to determine the masses of all planets in the system. 

Thanks to \textit{Kepler} and \textit{K2}, the population of planets known around young stars ($<$ 1 Gyr) has increased significantly (around 20\% of these planets have been discovered by this mission). Detecting and characterising these planets is important to test theories of planet formation and evolution. The current population of known planets around young stars comprises gaseous or icy planets and only one third of them have been detected in both transit and radial velocity, allowing a bulk composition to be calculated. So far, no rocky planets have been found in these young systems, although the large uncertainty on the mass of Kepler-289\,c \cite{schmitt14} does not rule out a rocky composition. \object{K2-233} is a young K3 dwarf star, with an estimated age of $360^{+490}_{-140}$~Myr \citep{david18}. Three transiting planets were detected based on \textit{K2} observations. We have now collected 124 HARPS radial velocity measurements to constrain their masses for the first time.

In this paper we present the characterisation of the properties of the detected planets in these two systems. We describe the observations in Section~\ref{sec:obs} and the methodology for the analysis in Section~\ref{sec:pastis}. The results are presented and discussed in Section~\ref{sec:discussion} and we provide conclusions in Section~\ref{sec:conclusions}

%________________________________________________________________
\section{Observations}
\label{sec:obs}

%-----------------------------------
\subsection{K2 photometry}
\label{sec:K2}

\begin{figure*}
\centering
\includegraphics[width=1\textwidth]{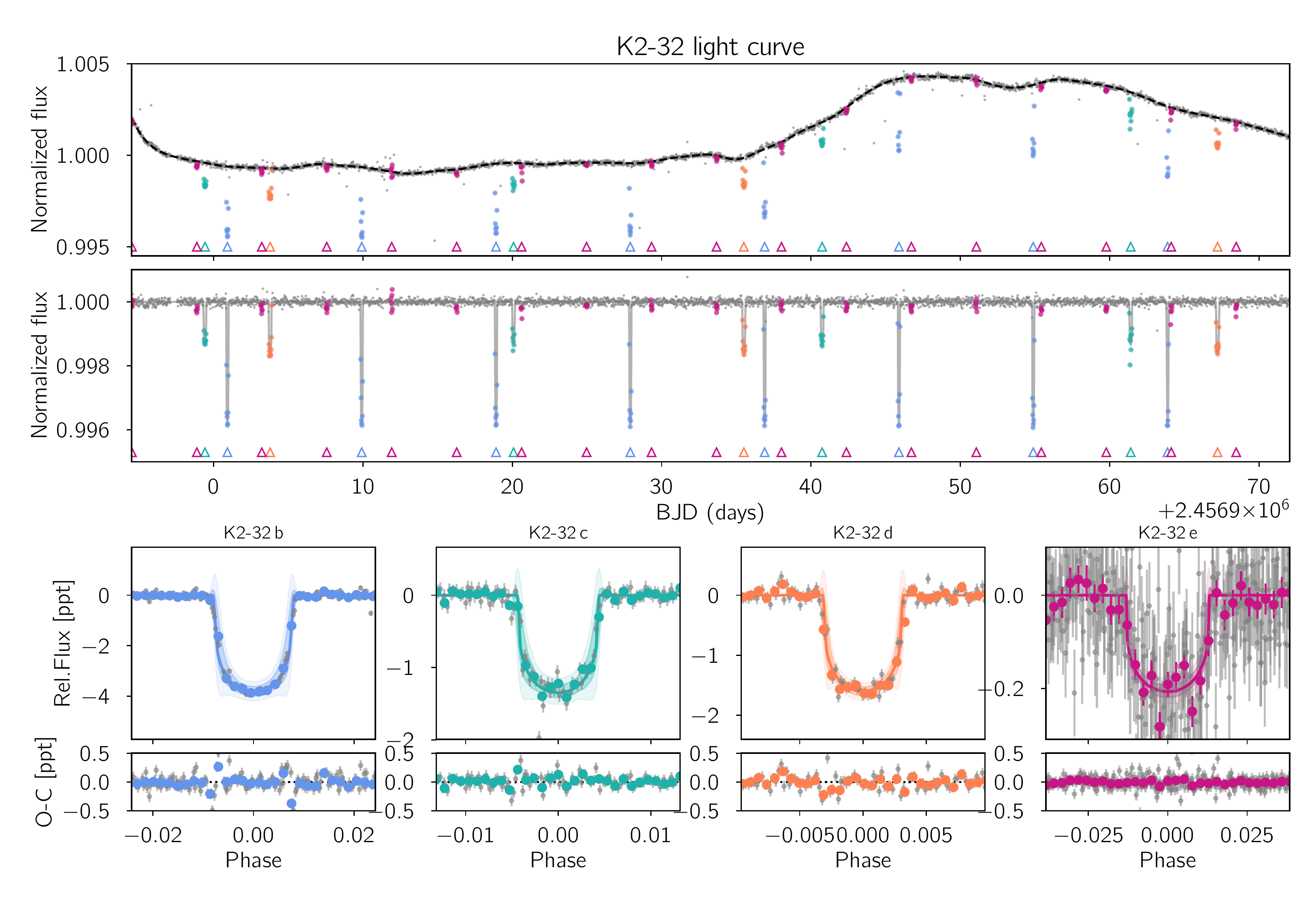} %/Users/lillo_box/00_projects/11__K2-32_K2-233/LC_RV_plots/
\caption{Photometric data and inferred model for K2-32. {Top:} K2 extracted photometry (grey dots, see Sect.~\ref{sec:K2}) and mean GP model (black dashed line). The transits for each individual planet are marked by upward triangles and colour-coded in the light curve.  {Middle:} Detrended light curve after removing the GP model (see Sect.~\ref{sec:K2}) and including the median transit model obtained with \texttt{PASTIS} for the 3 planets in the system (grey solid line). {Bottom:} Phase-folded lightcurves centred on the transit of each planet. The colour-coded symbols correspond to binned photometry with a bin size of  10\% of the transit duration. The inferred transit models are shown as solid lines. The shaded regions correspond to 68.7\%  (dark colour) and 95\% (light colour) confidence intervals.}
\label{fig:K2-32_data}
\end{figure*}

K2-32 was observed during Campaign 2 of the \textit{K2} mission for 77.5 days between 2014-Aug-23 and 2014-Nov-13, with a cadence of 29.4 minutes (3524 data points) and a median precision per data point of 71 parts per million (ppm), see Fig.~\ref{fig:K2-32_data} (upper panel). Three planets with periods 9 (b), 20.7 (c), and 31.7 (d) days were detected by \cite{sinukoff16} transiting this star during the \textit{K2} observations. The three planets were validated by the authors using \texttt{vespa} \citep{morton12}, with the caveat that K2-32c had a FPP~2.2\%, larger than the 1\% acceptable for validating purposes, but boosted by the multi-planet nature \citep{lissauer14} of the system the authors considered the planet as validated. By using the Transit Least Square (TLS) method, \cite{heller19} detected a fourth inner Earth-size planet  in the system (K2-32\,e, R$_e=1.01^{+0.10}_{-0.09}$~\rearth) transiting in the \textit{K2} data with  a period of $\sim4.3$ days, thus placing this multi-planetary system in a near 1:2:5:7 mean motion resonant chain. As mentioned in the discovery paper, despite some faint sources lying {inside the optimised photometric aperture}, the transits occur on the mentioned target and the companion sources are too faint to contaminate the planet parameters (see \citealt{sinukoff16}).  

K2-233 was observed by \textit{K2} in Campaign 15 of the \textit{K2} mission from 2017-Aug-23 to 2017-Nov-20, in a total time span of 76 days with 29.4-min cadence (3055 data points) and a median precision per datapoint of 58 ppm. No contaminant sources were found inside the \textit{K2} aperture. \cite{david18} detected three small planets (R$_p\approx 1.4, 1.3, 2.6$~\rearth) transiting this young star with periods 2.5 (planet b), 7.1 (planet c) and 24.2 (planet d) days. The planets were validated using the \texttt{vespa} \citep{morton12} algorithm.

The light curves were extracted by using the EPIC Variability Extraction and Removal for Exoplanet Science Targets \texttt{EVEREST} pipeline \citep{luger16,luger18} (see Tables \ref{tab:k2-32_lc} and \ref{tab:k2-233_lc}). We used the \texttt{EVEREST} light curves and not the \texttt{POLAR} ones \citep{barros16} as the systematics are slightly better corrected with the former. In the case of K2-32, this light curve was also used in the discovery paper to find the three external planets in the system and by \cite{heller19} to find the fourth planet (K2-32\,e). 
We first produced a transit mask to remove those epochs affected by the planet transits from the lightcurve. This was used to compute the \texttt{EVEREST} detrending and then, along with a mask of outliers identified by \texttt{EVEREST}, we used this time series of stellar flux to minimise the hyperparameters of a Squared Exponential kernel gaussian process (GP) using \texttt{george} \citep{george}.The resulting best-fit GP kernel was then applied to the full array of K2 epochs in order to predict and detrend the contribution of stellar variability during the masked transits and at outliers. {In the case of K2-233, the resulting long-term detrended light curve still showed a high-frequency correlated variability at around 6-hour periodicity potentially related to residuals of the 12-hour drift correction and thrusters firing to keep the pointing of the telescope \citep{howell14}. We removed this contribution through a spline filtering (with the planet transits masked), which reduced the photometric noise from 70 ppm to 40 ppm in K2-233, while preserving the transit shapes.}

\begin{figure*}
\centering
\includegraphics[width=1\textwidth]{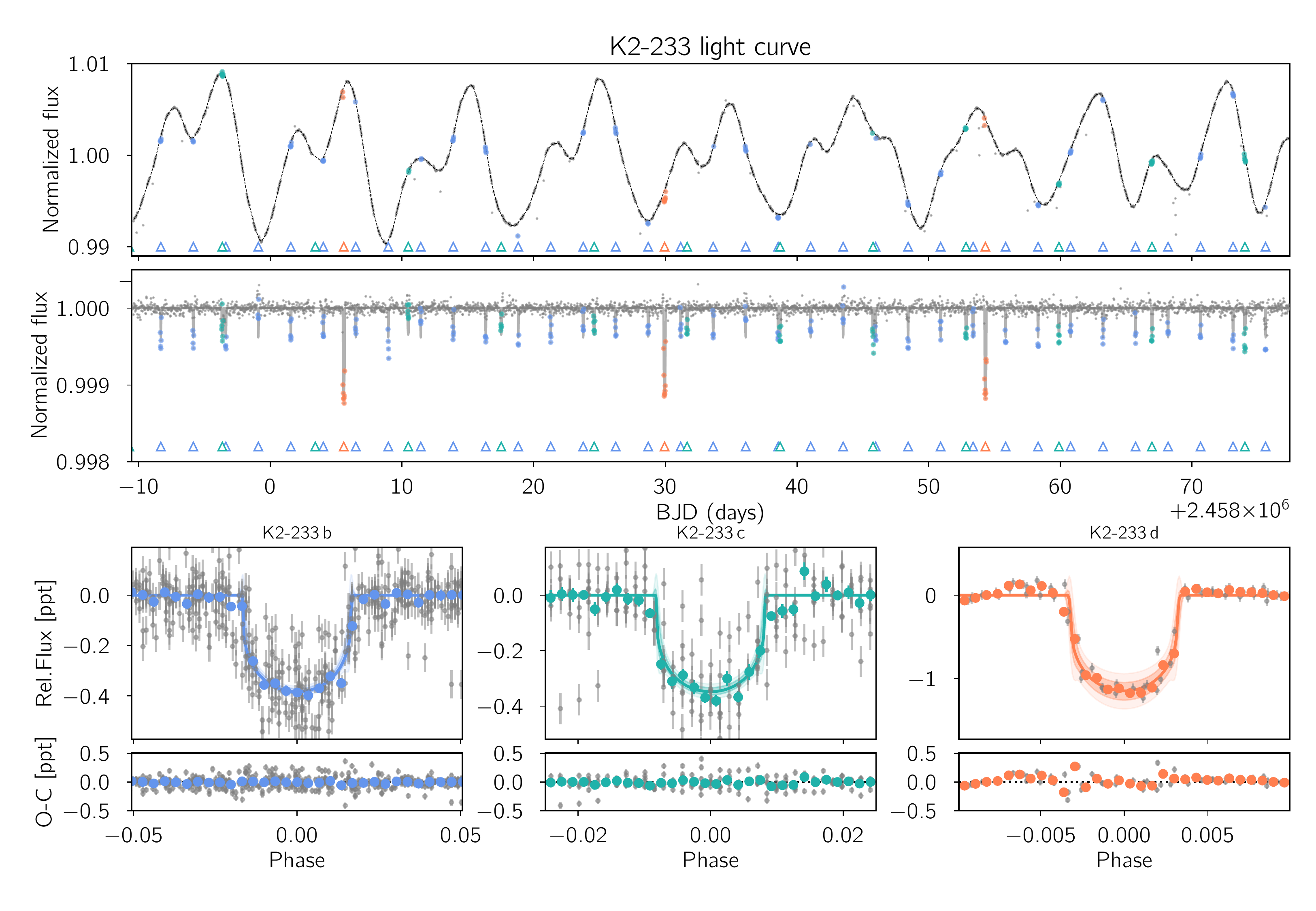} % /Users/lillo_box/00_projects/11__K2-32_K2-233/LC_RV_plots/
\caption{Photometric data and inferred model for K2-233. {Top:} K2 extracted photometry (grey dots, see Sect.~\ref{sec:K2}) and mean GP model (black dashed line). The transits for each individual planet are marked by upward triangles and colour-coded in the light curve.  {Middle:} Detrended light curve after removing the GP model {and the spline high-frequency filtering} (see Sect.~\ref{sec:K2}) and including the median transit model obtained with \texttt{PASTIS} for the 4 planets in the system (grey solid line). {Bottom:} Phase-folded lightcurves centred on the transit of each planet. The colour-coded symbols correspond to binned photometry with a bin size of  10\% of the transit duration. The inferred transit models are shown as solid lines. The shaded regions correspond to 68.7\%  (dark colour) and 95\% (light colour) confidence intervals.}
\label{fig:K2-233_data}
\end{figure*}

Figures \ref{fig:K2-32_data} and \ref{fig:K2-233_data} show the corresponding light curves at different stages for K2-32 and K2-233, respectively. For each system, the raw extracted light curve from \texttt{EVEREST} is shown in the upper panel, the detrended and normalised light curve after removing the mean GP model and the spline filtering is shown in the middle panel, and the phase-folded transit of each planet in the system is shown in the bottom panels.

%-----------------------------------
\subsection{Radial velocities}

In the case of K2-32, {by using 43 HARPS and 6 PFS observations}, \cite{dai16} could measure the mass of K2-32\,b to be $21.3\pm 5.9$~\mearth~. They could also place upper limits on the masses of the other two outer planets of M$_c<8.1$~\mearth\ and M$_d<35$~\mearth. \cite{petigura17} added  31 HIRES radial velocity measurements and found masses of $16.5\pm 2.7$~\mearth\ (planet b), $<12.1$\mearth\ (planet c), and $10.3\pm4.7$ (planet d), in all cases assuming circular orbits due to the lack of evidence for eccentric solutions. 

\begin{figure*}
\centering
\includegraphics[width=1\textwidth]{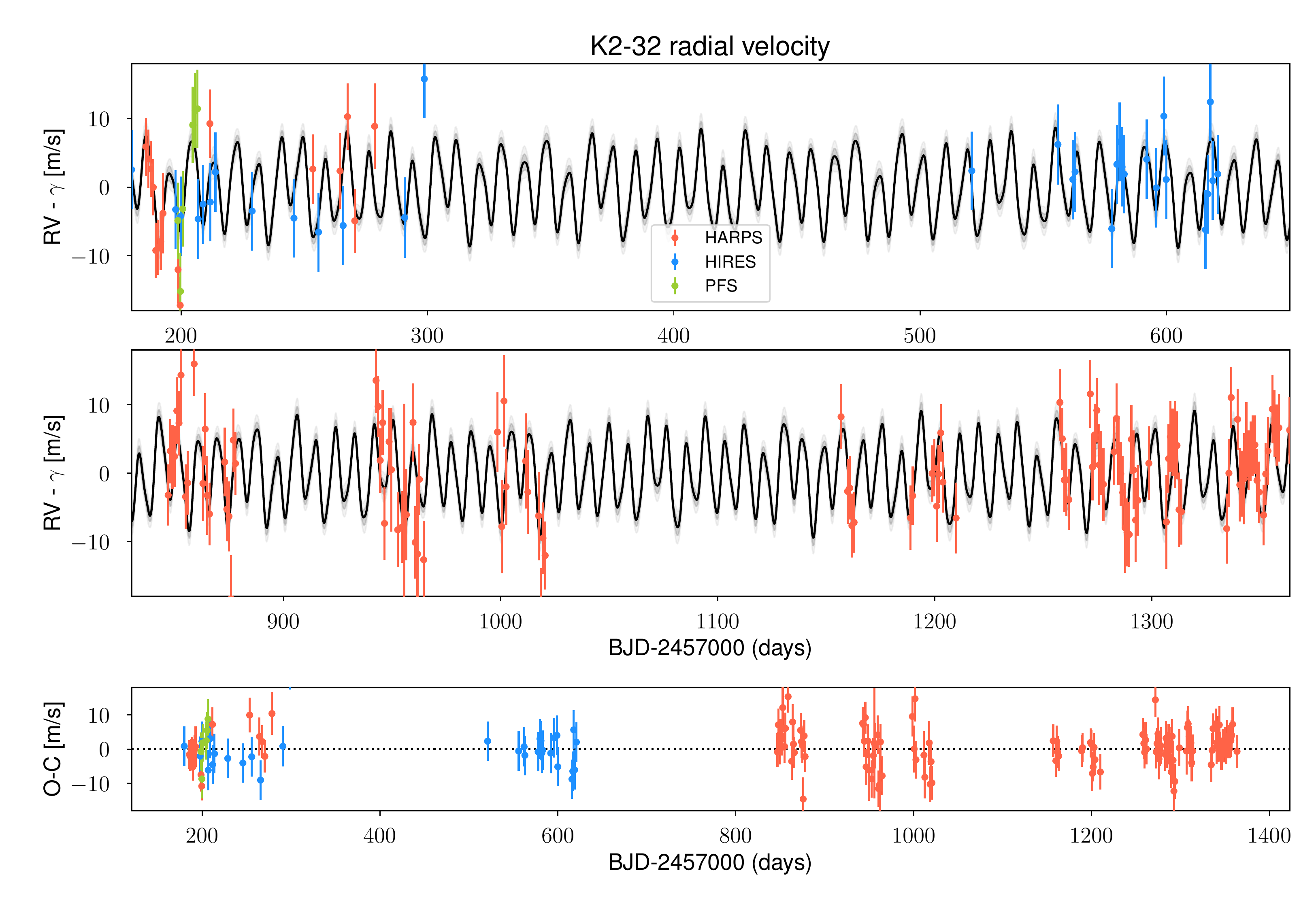} % /Users/lillo_box/00_projects/11__K2-32_K2-233/LC_RV_plots/
\caption{Radial velocity time series for the K2-32 planetary system. {Top and middle panels:} Full radial velocity dataset including observations from HARPS (red symbols), HIRES (blue symbols), and PFS (green symbols). The median model inferred with \texttt{PASTIS} including the four planets and the GP (see Section~\ref{sec:pastis}) is shown in black with 68.7\%  (dark grey) and 95\% (light grey) confidence intervals as shaded regions. {Bottom panel:} Residuals of the radial velocity modelling for the full dataset. {The white noise jitter term has been quadratically added to the radial velocity uncertainties in all panels.}}
\label{fig:K2-32_RV}
\end{figure*}

We have collected 199 new HARPS spectra during three semesters (Prog. ID: 198.C-0169, PI: A. Santerne). We have homogeneously reduced these data together with the 45 public spectra {(from Prog. ID 095.C-0718, PI: S. Albrecht; 43 of them published in \cite{dai16} and two additional with Julian dates 2457187.616519 and 2457192.682198)} in order to obtain an homogeneous HARPS radial velocity dataset by ensuring the same coordinates are used to compute the barycentric velocity correction. We cross-correlate the spectra with a binary G2 mask \citep{baranne96} and derive the radial velocities for each epoch by fitting a Gaussian function to the cross-correlation function. The final set {has a median radial velocity uncertainty of} 3.5 m/s. {In addition}, we have included in our analysis the 31 HIRES measurements from \cite{petigura17} and the 6 PFS radial velocities from \cite{dai16}. Altogether, our dataset comprises 279 radial velocities covering a total time span of 3.2 years. We performed a nightly binning of our HARPS radial velocities to correctly average the correlated high-frequency noise (mostly granulation), as it is known to bias the parameters derived, like the eccentricity \citep{hara19}. This can be done as the shortest planetary orbital period is 4.3 days and the vast majority of our data points obtained during the same night have time separations shorter than 4 hours (corresponding to $<$4\% of the orbital period).  The whole dataset is presented in Table~\ref{tab:k2-32_rv} and shown in Fig.~\ref{fig:K2-32_RV}.

\begin{figure*}
\centering
\includegraphics[width=1\textwidth]{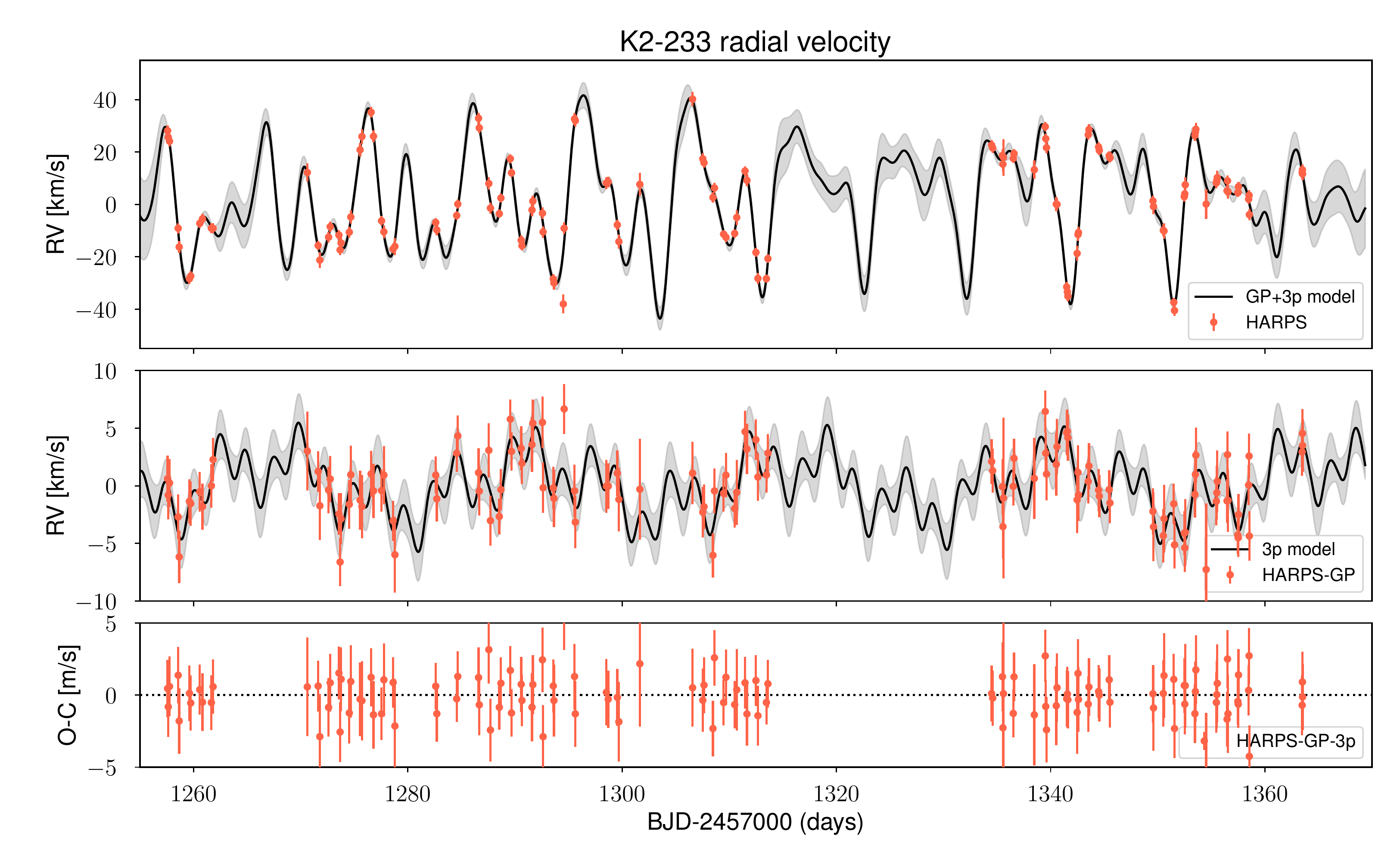} % /Users/lillo_box/00_projects/11__K2-32_K2-233/LC_RV_plots/
\caption{Radial velocity time series for the K2-233 planetary system. {Top panel:} Full HARPS radial velocity dataset. The median model inferred with \texttt{PASTIS} including the three planets and the GP (see Section~\ref{sec:pastis}) is shown in black with 68.7\% confidence intervals as grey shaded region. {Middle panel:} Radial velocity time series after removing the stellar activity contribution (i.e., the GP model), see Section~\ref{sec:pastis}. The full Keplerian model inferred for the three planets is shown in black with 68.7\%  (dark grey) and 95\% (light grey) confidence intervals as shaded regions. {Bottom panel:} Residuals of the radial velocity modelling.  {The white noise jitter term has been quadratically added to the radial velocity uncertainties in all panels.}}
\label{fig:K2-233_RV}
\end{figure*}

For K2-233, we have collected 124 HARPS spectra covering a 105-day timespan. The data reduction and radial velocity extraction was performed in a similar fashion as explained above. In this case, we obtain a median radial velocity {uncertainty} of 1.9 m/s. The radial velocities are presented in Table~\ref{tab:k2-233_rv} and shown in Fig.~\ref{fig:K2-233_RV}. {A clear large amplitude variation can be seen in the upper panel of Fig.~\ref{fig:K2-233_RV}, corresponding to the induced radial velocity modulated by the stellar activity of this star}.  

\begin{figure*}
\centering
\includegraphics[width=0.48\textwidth]{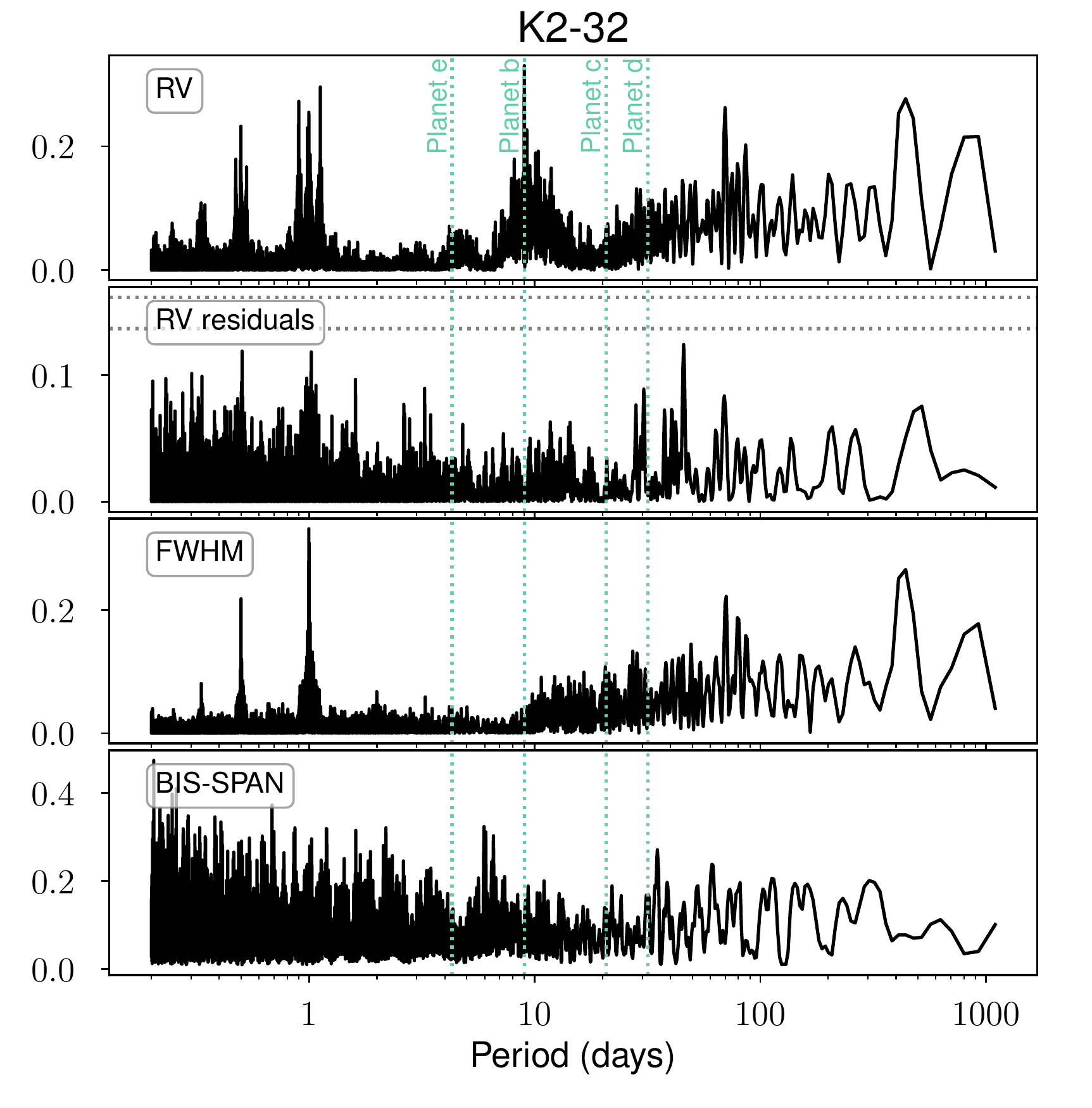} %/Users/lillo_box/00_projects/11__K2-32_K2-233/periodograms/
\includegraphics[width=0.48\textwidth]{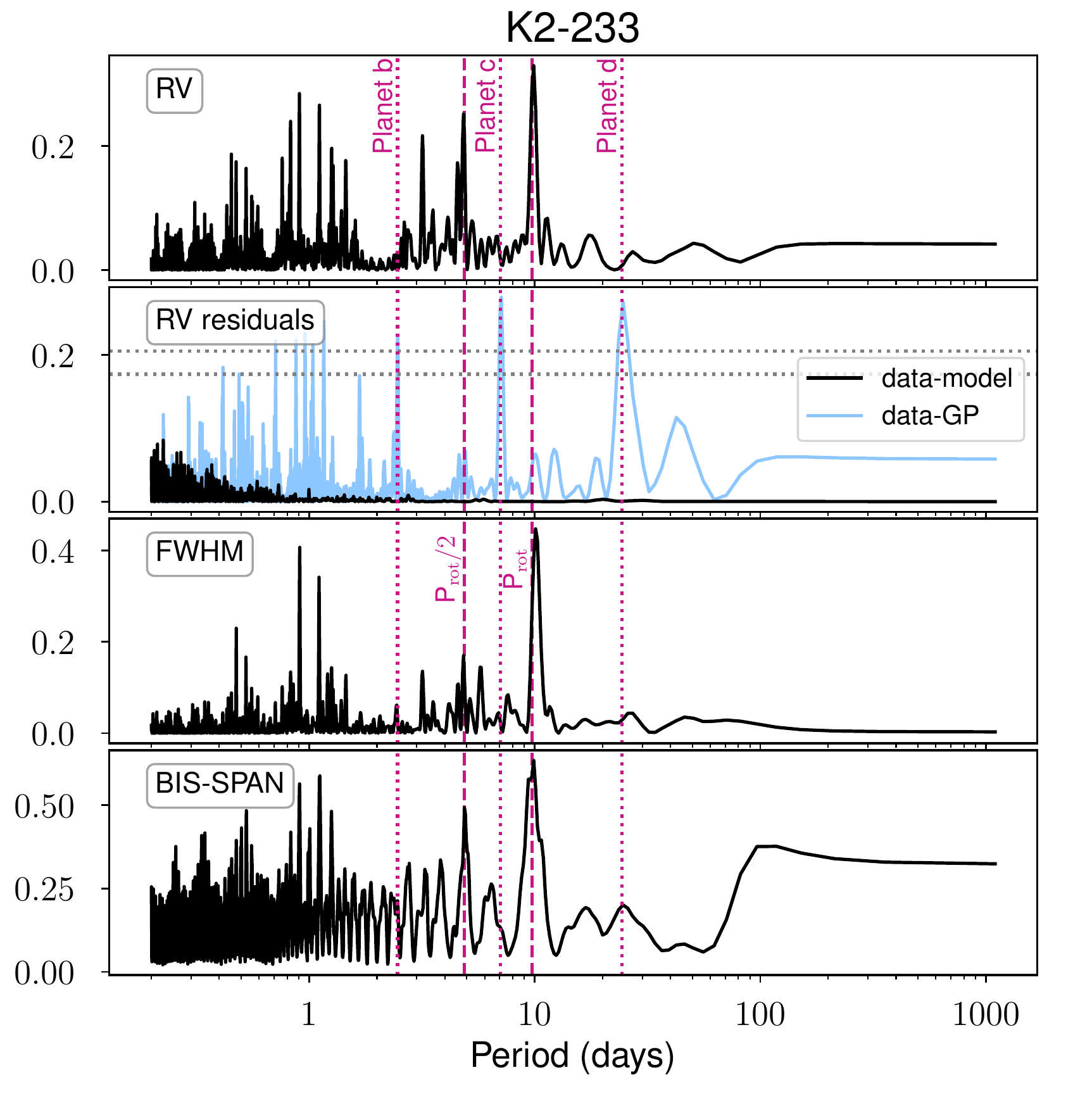} %/Users/lillo_box/00_projects/11__K2-32_K2-233/periodograms/
\caption{{Lomb-Scargle periodogram of the HARPS radial velocity datasets and activity indicators from K2-32 (left panels) and K2-233 (right panels). For each system we show, from top to bottom, the periodogram corresponding to: the radial velocity, the residuals after subtracting the radial velocity model (for K2-233 we also show the residuals of the GP model) and the 1\% and 0.1\% false alarm probabilities as horizontal dotted lines, the full-width at half maximum of the cross-correlation function and the bisector span. The planet periods are marked as dotted lines and the rotation period of K2-233 and its first harmonic are marked as dashed lines.}}
\label{fig:per}
\end{figure*}

{The periodogram of both radial velocity datasets is shown in Fig.~\ref{fig:per} together with the activity indicators corresponding to the full-width at half-maximum (FWHM) and the bisector span. In the case of K2-233, the periodogram is dominated by the stellar activity of the star and none of the planet signals are visible.} 

%________________________________________________________________
\section{Spectroscopic stellar parameters}
\label{sec:stellar_params}
We used \texttt{ARES}+\texttt{MOOG} \citep[\texttt{ARES} v2 and \texttt{MOOG2014}; for details, see][]{sousa14} to derive spectroscopic parameters of the two host stars studied in this work. The spectral analysis is based on the excitation and ionisation balance of iron abundance. The equivalent widths of the iron lines were consistently measured with the \texttt{ARES} code \citep[][]{Sousa-2007, sousa15} and the abundances are derived in local thermodynamic equilibrium (LTE) with the \texttt{MOOG} code \citep[][]{Sneden-1973}. For this step we use a grid of Kurucz ATLAS9 plane-parallel model atmospheres \citep[][]{Kurucz-1993}. The list of iron lines is taken from \citet[][]{Sousa-2007} except for K2-233 (with effective temperature below 5200 K) where we use instead a more adequate line list for cooler stars \citep[][]{Tsantaki-2013}. This method has been applied in our previous spectroscopic studies of planet-hosts stars which are all compiled in the Sweet-CAT catalogue \citep[][]{santos13, Sousa-2018}. For the derivation of chemical abundances of the refractory elements we used the same tools and models of atmospheres as for the derivation of the stellar parameters. The list of spectral lines and the method is described in \cite{adibekyan12,adibekyan15}. 

The resulting parameters from this analysis are presented in Table~\ref{tab:specparams}. {For K2-32, the effective temperature ($5273\pm25$~K) and surface gravity ($4.359\pm0.048$~dex) values are consistent within 1$\sigma$ with those published by \cite{dai16} and slightly lower than those from \cite{mayo18} who used data from the TRES spectrograph and compared them to a library of stellar spectra from \cite{buchhave12}. For K2-233,  we obtained different values in our spectroscopic analysis than those published by \cite{david18}. In our case, we estimate a slightly cooler effective temperature of $T_{\rm eff}=4796\pm66$~K (against the $T_{\rm eff}=4950\pm100$~K found by \citealt{david18}, still compatible within 1$\sigma$) but a relevant difference in the surface gravity, with $4.30\pm0.19$ found in this work and $4.71\pm0.1$ found in \cite{david18}, so around a $2\sigma$ discrepancy.  We note that in this case if we use the correction for the spectroscopic surface gravity derived from asteroseismology and presented in \cite{mortier14}, we obtain $4.53\pm0.22$, which becomes compatible with previous studies within 1$\sigma$. We used our spectroscopic estimates as Gaussian priors in the subsequent \pastis{} analysis.} 

\input{Table_SpectroscopicParams.tex}

%________________________________________________________________
\section{Joint analysis with \pastis{}}
\label{sec:pastis}

The photometric and radial velocity data were jointly analysed with the Spectral Energy Distribution (SED) in a Bayesian framework using the \pastis{} software (see \citealt{diaz14b}, and applications of this software in \citealt{barros17}, \citealt{santerne18} or \citealt{lam18}). The magnitudes for the SED were extracted from the American Association of Variable Star Observers Photometric All-Sky Survey (APASS, \citealt{henden15}) archive in the optical, and from the Two-Micron All-Sky Survey (2MASS, \citealt{munari14}) and the Wide-field Infrared Survey Explorer (AllWISE, \citealt{cutri14}) archives in the near-infrared. The complete list of magnitudes is reported in Table~\ref{tab:stellarparam1}. The SED was modelled with the BT-Settl stellar atmospheric models \citep{allard12}. {The K2-32 masses and orbital parameters were estimated by modelling the radial velocities with Keplerian orbit models}. {No Gaussian process was needed in this case} since K2-32 is a quiet star at the level of precision we probed. For K2-233, we used both Keplerian orbit models and a Gaussian process regression with a quasi-periodic kernel {to simultaneously account for the correlated noise induced by the activity of this young and active star}. We used the following form for the GP kernel:

\begin{equation}
\begin{split}
    k(t_i, t_j) = &~A^2 ~\exp\left[ -\frac{1}{2} \left( \frac{t_i-t_j}{\lambda_1} \right) ^2 - \frac{2}{\lambda_2^2} \sin^2 \left( \frac{\pi \left| t_i-t_j \right| }{P_{rot}} \right) \right] + \\
    	& +~\delta_{ij} \sqrt{\sigma_i^2 + \sigma_{J}^2},
\end{split}
\end{equation}

\noindent where $A$ corresponds to the modulation amplitude, P$_{\rm rot}$ to the stellar rotation period, $\lambda_1$ to the timescale of the correlations decay, $\lambda_2$ is a measure of the relative importance between the periodic and the decaying components, { and $\sigma_{J}$ is the radial velocity jitter}. 

To model the photometry, we used the JKT Eclipsing Binary Orbit Program (\texttt{JKTEBOP}, \citealt{southworth08}) with an oversampling factor of 30 to account for the long integration time of \textit{Kepler} \citep{kipping10}. { For K2-32, t}he central star was modelled with the Dartmouth evolution tracks \citep{dotter08}, taking into account the asterodensity profiling \citep{kipping14}. We {also tried using the PARSEC} evolution tracks \citep{bressan12}, with completely consistent results. {For K2-233, as the star is relatively young and to avoid biaising the determination of the stellar mass and radius, we used the constraints on the stellar age from \citet{david18} along with the PARSEC evolution tracks which start earlier than the Dartmouth tracks}. Finally, the limb darkening coefficients were taken from \citet{claret11}, using the stellar parameters at each step of the analysis.

\texttt{PASTIS} runs Markov Chain Monte Carlo (MCMC) to explore the posterior distributions of the various parameters. We ran 40 chains with $6 \times 10^5$ iterations for each system. {The chains were initialised} randomly from the joint prior distribution. For the priors, we used normal distributions centred on the results from the spectral analysis for the stellar effective temperature, surface gravity, and metallicity  of K2-32. {For K2-233, we used a uniform prior on the age between 100 Myr and 1 Gyr, a broad normal distribution around the initial stellar mass derived by \cite{david18}, and a normal distribution on the iron abundance inferred in our spectroscopic analysis}. For the systemic distance to Earth, we used a normal distribution centred on the \textit{Gaia} data release 2 from \cite{gaia18} {(parallax of $\pi = 6.313 \pm 0.052$~mas for K2-32 and $\pi = 14.785 \pm 0.061$~mas for K2-233)} , taking into account the correction {($\delta\pi=0.054$~mas)} from \citet{schonrich19}. For the planetary period and transit epoch, we used a normal distribution centred on the ephemeris from the transit identification {and with a width larger than the published uncertainties}. For the eccentricity, we used a truncated normal distribution with a width of 0.083, {valid for small planets in multi-transiting systems} \citep{vaneylen19}. Finally, for the planetary orbital inclination, we used a sine prior {(a uniform prior in $\sin{i}$)}. The priors of the remaining parameters are set to uniform distributions (see Table~\ref{tab:K32_MCMC} and \ref{tab:K233_MCMC}).

\begin{figure*}
\centering
\includegraphics[width=1\textwidth]{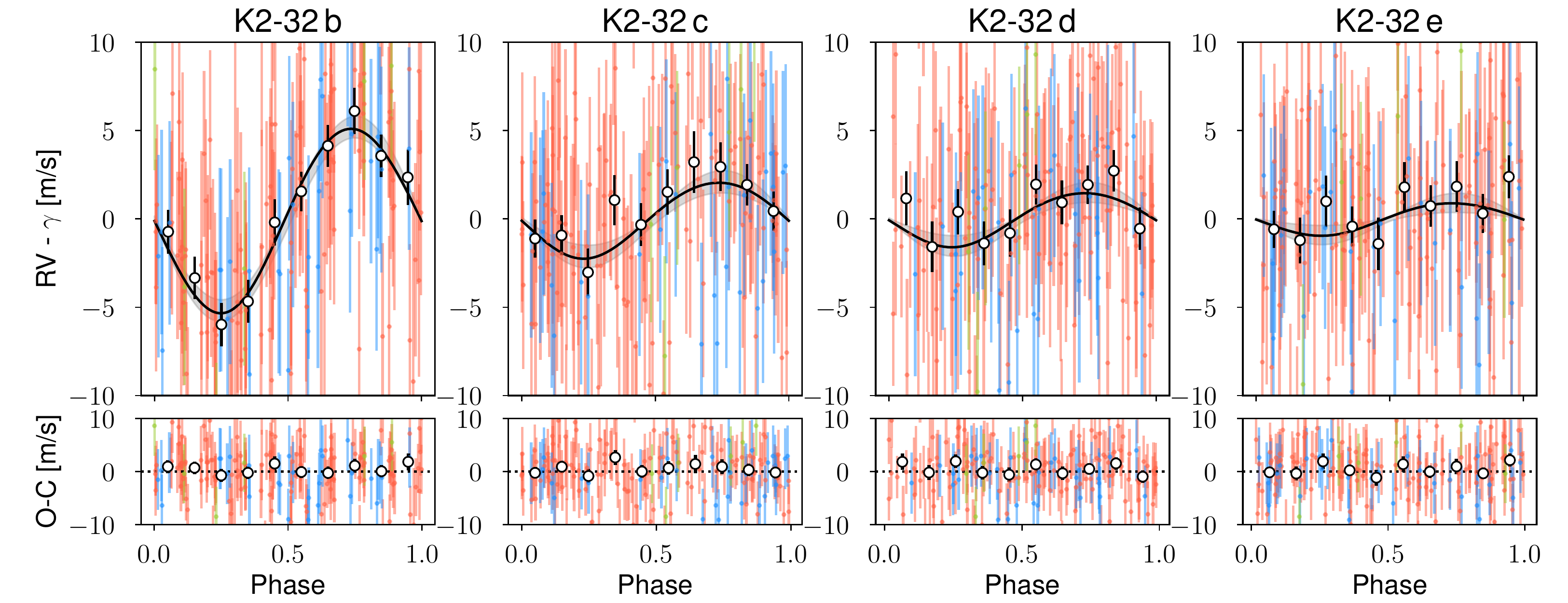} %/Users/lillo_box/00_projects/11__K2-32_K2-233/LC_RV_plots/
\caption{{Phase-folded radial velocity curves for the four planets in the K2-32 planetary system. The different colour correspond to HARPS (red symbols), HIRES (blue symbols), and PFS (green symbols) radial velocities. The big black open circles correspond to average radial velocities in 10\% phase bins. The black solid line shows the median model inferred by \texttt{PASTIS} and the shaded regions correspond to 68.7\%  (dark grey) and 95\% (light grey) confidence intervals. Residuals of the model are shown in the bottom panels.}}
\label{fig:K2-32_RVphase}
\end{figure*}

\begin{figure*}
\centering
\includegraphics[width=1\textwidth]{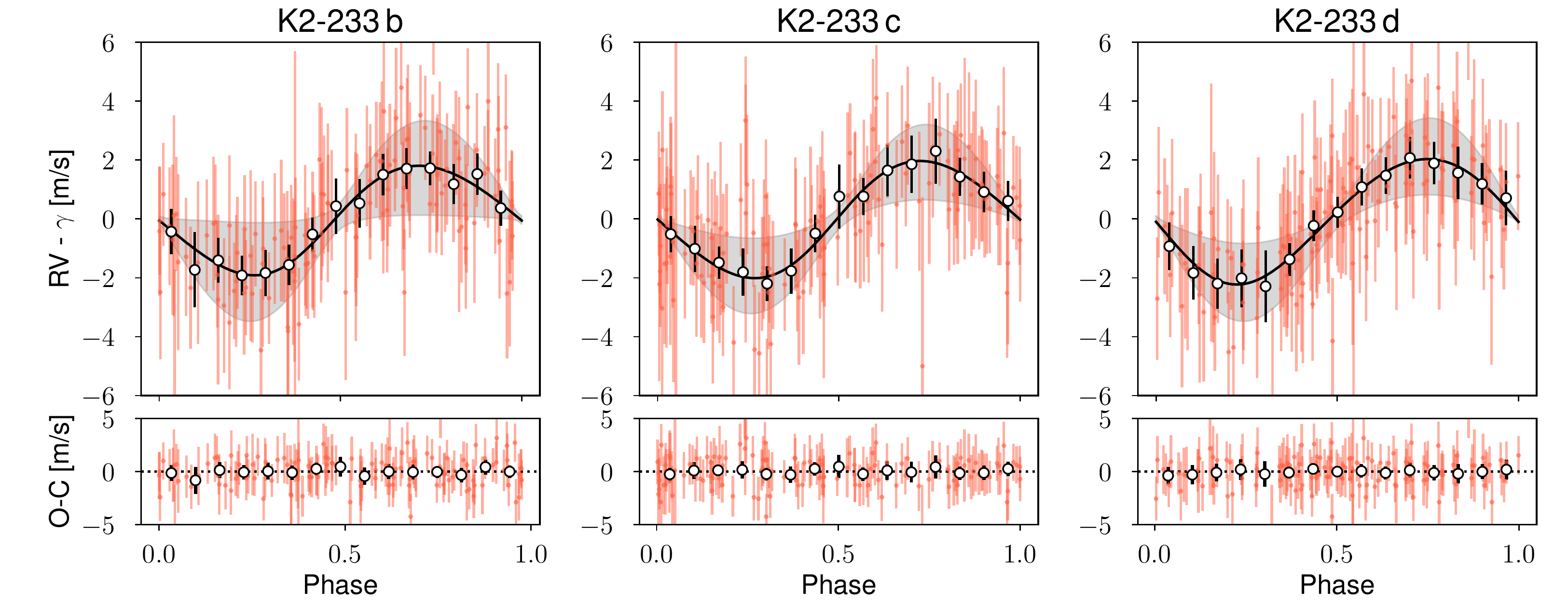} %/Users/lillo_box/00_projects/11__K2-32_K2-233/LC_RV_plots/
\caption{{Phase-folded HARPS radial velocity curves for the four planets in the K2-233 planetary system. The big black open circles correspond to average radial velocities in 10\% phase bins. The black solid line shows the median model inferred by \texttt{PASTIS} and the shaded regions correspond to 68.7\%  (dark grey) and 95\% (light grey) confidence intervals. Residuals of the model are shown in the bottom panels.}}
\label{fig:K2-233_RVphase}
\end{figure*}

To improve the convergence and the number of samples in the posterior distributions, we re-ran a full analysis for both systems, starting from the best values of the first analysis. In both cases, the convergence of the chains was checked using a Kolmogorov-Smirnov test \citep{brooks03}. The burn-in of each chain {(determined as described in \citealt{diaz14b})} was then removed and they were merged to derive proper credible intervals for each parameter. The full list of priors and posteriors is reported in Tables \ref{tab:K32_MCMC} and \ref{tab:K233_MCMC}. {A summary of the derived stellar properties is shown in Table~\ref{tab:stellarpars} and the estimated and derived planet properties are summarised in Tables~\ref{tab:k32sum} (K2-32) and \ref{tab:k233sum} (K2-233)}. {The final light curve model and individual planet contributions are presented in the middle and bottom panels of Fig.~\ref{fig:K2-32_data} (for K2-32) and Fig.~\ref{fig:K2-233_data} (for  K2-233). The full radial velocity dataset and model (including confidence intervals) is shown in Figs.~\ref{fig:K2-32_RV} (K2-32) and \ref{fig:K2-233_RV} (K2-233). The isolated radial velocity signal of each planet (after removing the GP and contribution from the other planets) is shown in Figs.~\ref{fig:K2-32_RVphase} (K2-32) and \ref{fig:K2-233_RVphase} (K2-233)}.
In the second panel of Fig~\ref{fig:per} we show the Lomb-Scargle periodogram of the residuals of the radial velocities after removing the planets and GP models. {No significant periodicities are found in either system with the current dataset}.

\input{Table_Phot_mags.tex}
\input{Table_Stellar_parameters_summary.tex}
\input{Table_K2-32_summary_table.tex}

\input{Table_K2-233_summary_table.tex}

%________________________________________________________________
\section{Discussion}
\label{sec:discussion}

\begin{figure}
\centering
\includegraphics[width=0.5\textwidth]{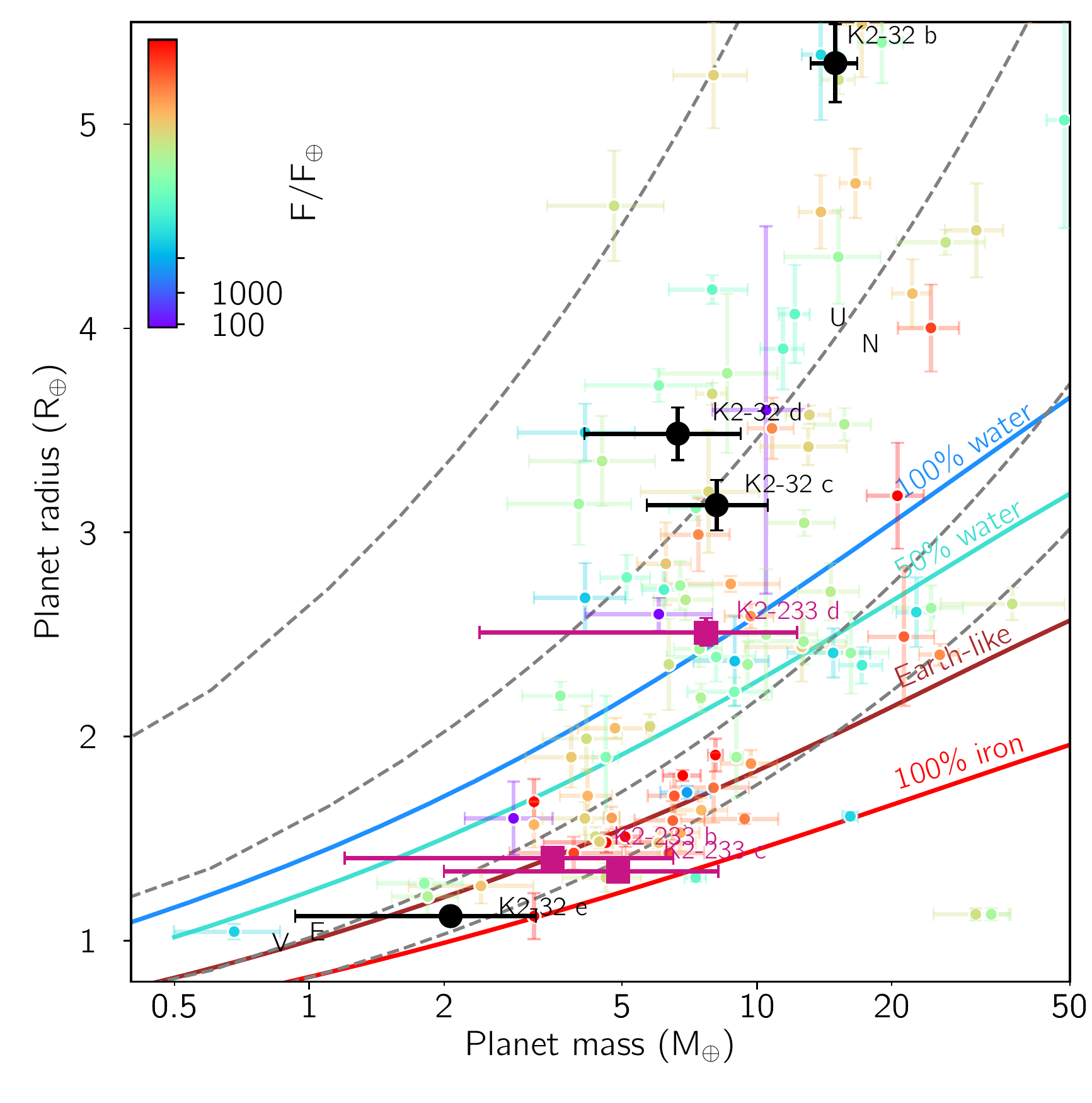} % /Users/lillo_box/11_MyPapers/0B_00_K2-32_K2-233__PlanetarySystems/plots/
\caption{Mass-Radius diagram for planets with masses less than 50~\Mearth. The colour code of the symbols shows the insolation relative to the Earth. The K2-32 system planets are shown as black circles and the K2-233 planets with purple square symbols. {Iso-density lines corresponding to different compositions from \cite{zeng19} are shown as solid traces while the dashed lines correspond to iso-densities of 0.3, 1.33, 5.3, 10 g$\cdot$cm$^{-3}$ (from top to bottom)}. }
\label{fig:mass-radius}
\end{figure}

%--------------------------
\subsection{K2-32: diverse compositions in a 4-planet system}

\begin{figure}
\centering
\includegraphics[width=0.5\textwidth]{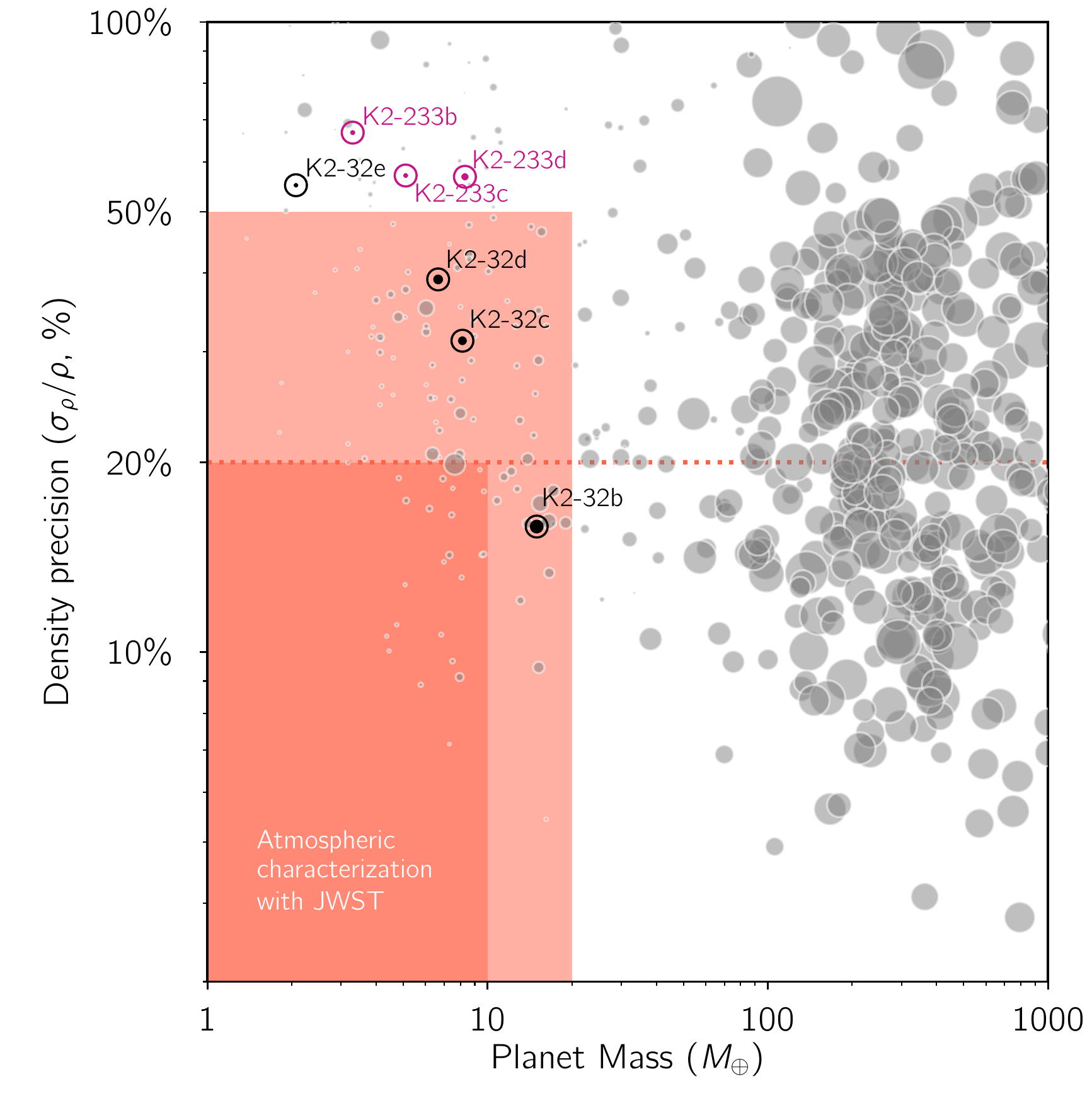}
\caption{Density precision versus mass for the known planets. The symbol size of all planets is scaled with their size. Planets studied in this work are marked by black (K2-32) and purple (K2-233) filled symbols and are highlighted by big open circles. The two shaded regions mark the optimal region for atmospheric characterisation with JWST for Neptune-size (light region) and rocky (dark region) worlds. }
\label{fig:period-dens}
\end{figure}

The analysis of the new 199 HARPS radial velocity measurements added to the archival data on this source has allowed us to characterise the planetary masses and densities of the four planets in this system (see Fig.~\ref{fig:mass-radius}, left panel). We determine a mass of $15.0^{+1.8}_{-1.7}$~\Mearth\ for K2-32\,b, {smaller} than the previously reported value in \cite{dai16} ($21.1\pm5.9$~\Mearth) and three times more precise, from $\sigma_{M_{\rm b}}/M_{\rm  b}=$~28\% to 12\%. Our values are consistent within 1$\sigma$ with the mass estimated by \cite{petigura17} although slightly more precise (from 16\% to 12\%). The analysis presented in this paper now translates into a precision on the bulk density of the planet of 18\% ($0.55\pm0.1$~\gcm3), which converts this gaseous planet into a good candidate for further atmospheric characterisation. We can also confirm for the first time the planets $c$ and $d$ with more than 99\% confidence in the determination of their mass, $8.1\pm2.4$~\Mearth\ and $6.7\pm2.5$~\Mearth\ respectively, with corresponding bulk densities of $\rho_c=1.43^{+0.48}_{-0.45}$~\gcm3 and $\rho_d=0.87^{+0.35}_{-0.34}$~\gcm3. These values place the two outer planets in the mini-Neptune regime. Finally, we provide a marginal detection of the inner small planet K2-32\,e, with a mass of $2.1^{+1.3}_{-1.1}$~\Mearth\ and a 95\% upper limit of 4.32$~\Mearth$, corresponding to a bulk density of $\rho_e=6.3^{+4.1}_{-3.5}$~\gcm3, placing this planet in the Earth-like rocky-composition regime. {These estimates now reduce the relative {uncertainty} on the bulk density of the three outer planets in the system below 40\%. Specially interesting is the case of K2-32\,b, where we obtain a relative uncertainty better than 20\% (see Fig.~\ref{fig:period-dens}),} critical for subsequent atmospheric studies with the \textit{James Webb} Space Telescope (JWST).  

The architecture of {this system somewhat resembles a scaled-down version} of the Solar System, with an inner rocky {planet} followed by a gaseous giant (sub-Saturn density) and two external {mini-Neptunes}. Its structure is also similar to the three inner planets of WASP-47 \citep{hellier12,becker15}. Interestingly, the four planets in K2-32 are in a near-resonant chain 1:2:5:7, being one of the few multi-planetary systems with more than 2 planets in this configuration. The determination of their masses also allows us to determine their mutual separations in terms of the Hill {radius} ($\Delta$). \cite{weiss18} found that adjacent planet pairs in multi-planet systems tend to be separated by a mean value of 20 Hill {radii} (the \textit{peas-in-a-pod} hypothesis). In the case of K2-32, we find these mutual separations to be 18, 18, and 11 respectively from inner pairs to outer pairs. It is relevant to highlight that only the 10\% of planet pairs in multi-planetary systems with four or more planets show $\Delta<11$. However, as noted by {Weiss et al.}, {this might be caused by a poorly estimated} mass from mass-radius relations. Indeed, in the case of Kepler-11, where all masses are known, the mutual separations are all $\Delta<10$. The precise determination of the masses in these systems shows that more dedicated Doppler programs on systems with four or more planets are needed to extract conclusions about the \textit{peas-in-a-pod} hypothesis.

\begin{figure}
\resizebox{\hsize}{!}{\includegraphics{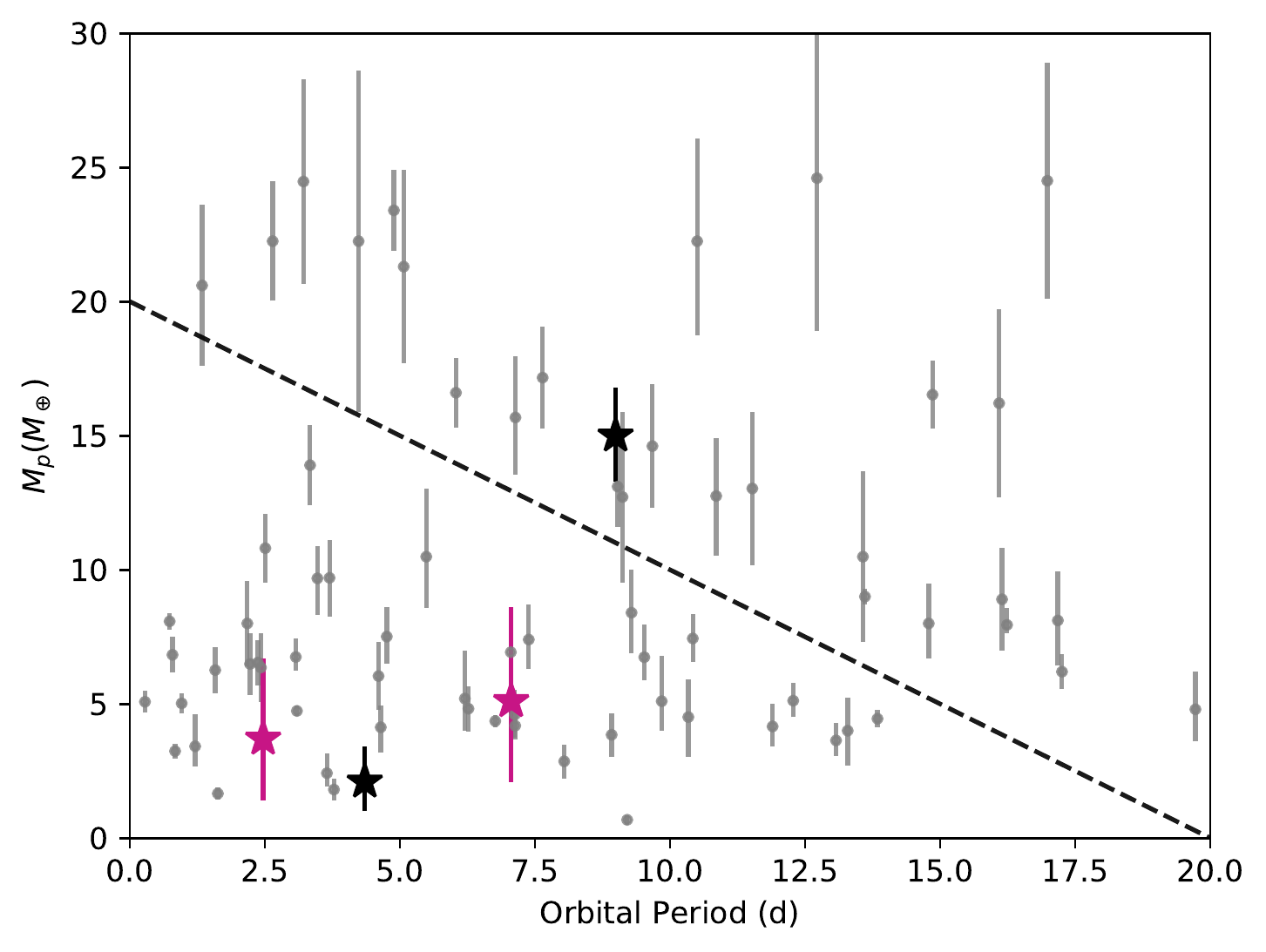}}
\caption{The mass-period gap from  \cite{armstrong19} (grey dots) with our results for K2-32\,$e$ and $b$ (black symbols) and K2-233\,$b$ and $c$ overplotted (violet symbols).}
\label{fig:mpgap}
\end{figure}

We consider our mass measurements in the context of the mass-period gap published in \cite{armstrong19} in Fig. \ref{fig:mpgap}. The mass-period gap may arise due to tidal interactions with the host star or interactions between the planet and protoplanetary disk. Of the four considered planets which have orbital periods shorter than 20~days {(K2-32\,b, K2-32\,e, K2-233\,b, and K2-233,\,c)}, all lie outside the gap, with K2-32 one of a handful of systems whose planets straddle the gap. The underlying cause of this structure in the planet distribution remains uncertain, but additional planetary mass measurements will help to illuminate the formation and evolution processes {that} sculpt the planet population in this region of parameter space.

%--------------------------
\subsection{K2-233: the first mass constraints for young rocky planets}

The HARPS radial velocities obtained for K2-233 in this program have allowed us to put constraints on the planet masses in this young planetary system. {We find $M_b=3.3^{+3.3}_{-2.2}$\Mearth{}, $M_c=5.1^{+3.2}_{-2.9}$\Mearth{}, and $M_d=8.3^{+5.2}_{-4.7}$\Mearth{}}. While K2-233\,d can now be {identified as a most likely sub-Neptune-like gaseous planet}, the other two inner components of the system (K2-233\,b and K2-233\,c) have densities compatible with rocky worlds {and lie} on the left-hand side of the  radius valley {located around 1.8~\Rearth\ } \citep{fulton17}.

\begin{figure*}
\centering
\includegraphics[width=0.48\textwidth]{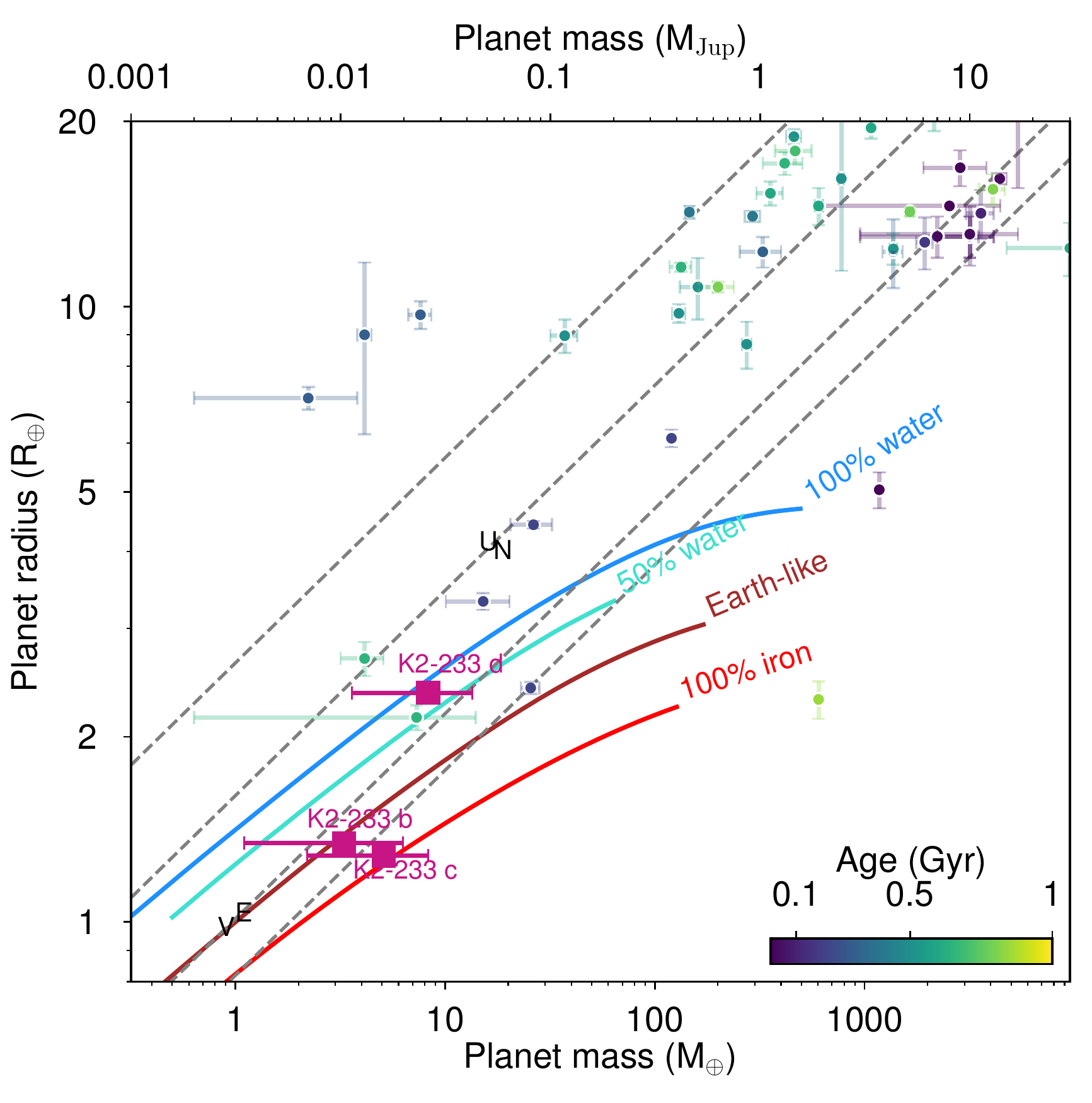}
\includegraphics[width=0.48\textwidth]{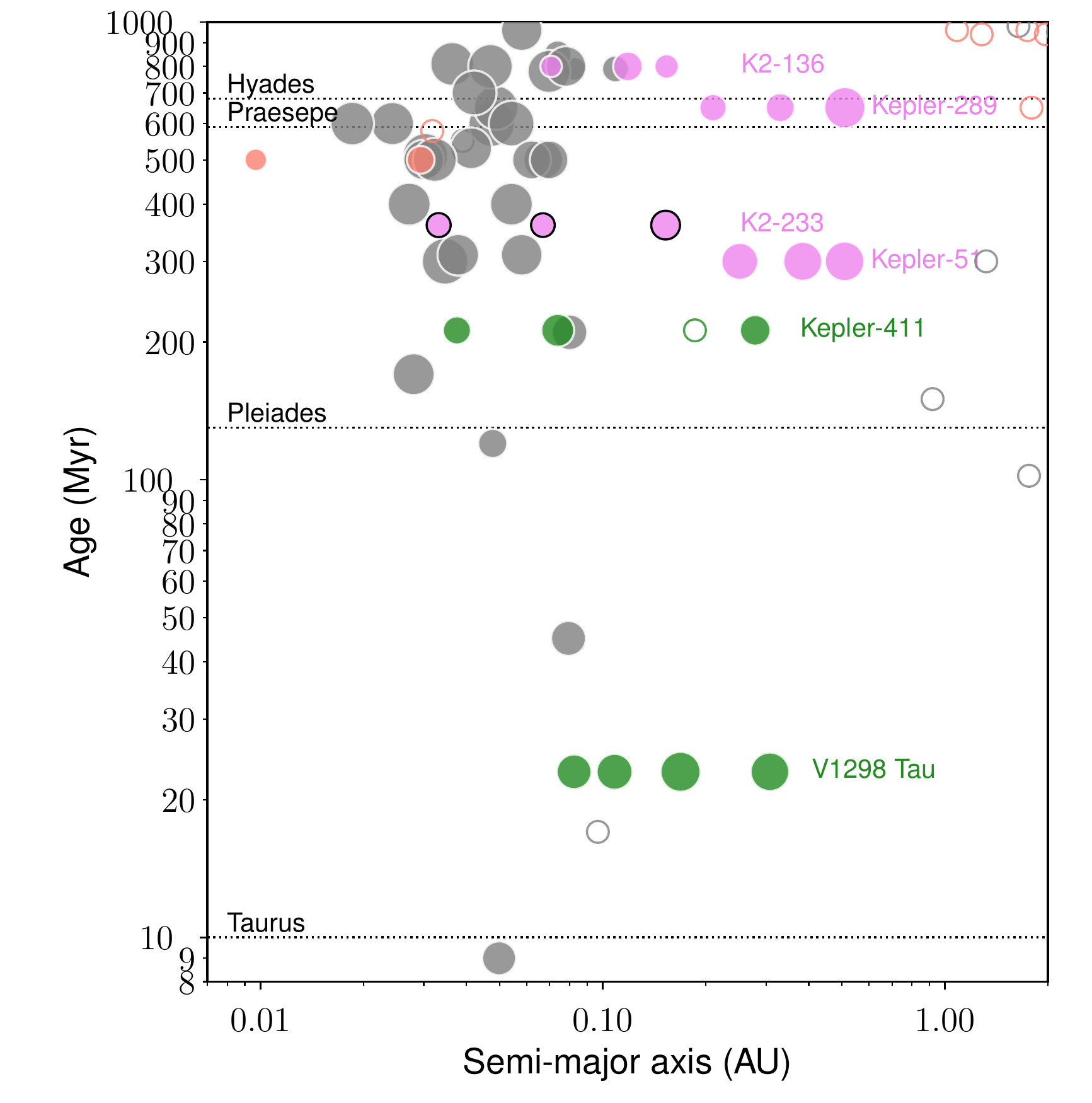}
\caption{ {Left:} Mass-Radius diagram for planets around stars with estimated ages below 1 Gyr. The colour-code represents the estimated median stellar age and the K2-233 planets are marked as violet square symbols. {Iso-density lines corresponding to different compositions from \cite{zeng19} are shown as solid traces while the dashed lines correspond to iso-densities of 0.3, 1.33, 5.3, 10 g$\cdot$cm$^{-3}$ (from top to bottom)}. {Right:} Planetary systems with ages younger than 1 Gyr against the semi-major axis of the planets. We show systems with one (grey symbols), two (red), three (pink), and four (green) known planets. Symbol sizes correspond to the planet size and non-transiting planets are represented as open circles. The young systems with three or more planets known (including K2-233) are also labeled. }
\label{fig:mass-radius_young}
\end{figure*}

The mass measurement of these two planets represents the first {mass} and bulk density estimation of a rocky world around a host star younger than 1 Gyr. These two planets are indeed the only ones in the {smaller mode of the radius distribution} in the sample of young planets (see Fig.~\ref{fig:mass-radius_young}). Two other systems share similar properties to K2-233. First, the young (212 Myr-old) four-planet system Kepler-411 has a similar architecture as K2-233, with the inner three planets having the same semi-major axis values as K2-233\,b,\,c,\,d, but being bigger and more massive (25, 26 and 11 \Mearth; 2.4, 4.4 \Rearth, with Kepler-411\,e non transiting). The star is also a K3V star, similar to the K2V spectral-type of K2-233. Hence, these two systems represent an ideal laboratory to study the evolution of multi-planetary systems {at early stages after the planet formation} ($212 \pm31$ Myr for Kepler-411 and $360^{+490}_{-140}$ Myr for K2-233). {On the other hand}, the Kepler-289 three-planet system orbits a slightly older (650 $\pm$ 440 Myr) solar-type star. In this case, the three planets are located further away, at 0.21, 0.33 and 0.51 AU. The system is indeed particularly similar to Kepler-51, also a solar-like star orbited by three planets at 0.25, 0.38 and 0.51 AU and with an estimated age of 310 Myr, {albeit} with large uncertainties. In this case, the planets are particularly inflated. K2-233 belongs to this small but growing family of young multi-planetary systems ($N_p>2$) and for the first time reaches the rocky-composition regime. 

The masses measured in this work allow us to establish the mutual separation of the adjacent planet pairs in the system in terms of the Hill radius. We can now precisely measure these values thanks to the mass measurements. {We find mutual separations of $30.2^{+6.1}_{-3.6}$ and $30.5^{+5.9}_{-3.6}$ mutual Hill {radii} for the two adjacent planet pairs}. We thus confirm the anomalously large separation between the planets of this system as compared to other three-planet systems, which tend to be more closely packed with mutual separations around 20 Hill radii \citep{weiss18}. 
Additional planets might also be present in the system but the current data does not allow their detection. We have tested their presence by analysing the HARPS RVs assuming a 3-planet model ($\mathcal{M}_{\rm 3p}$), a 4-planet model with a fourth planet between planets $b$ and $c$ ($\mathcal{M}_{\rm 4pA}$), a  4-planet model with a fourth planet between planets $c$ and $d$ ($\mathcal{M}_{\rm 4pB}$) and a 5-planet model with a planet between $b$ and $c$ and another planet between $c$ and $d$ ($\mathcal{M}_{\rm 5p}$). We have computed the evidence of these four models and we find the 3-planet model to have the largest evidence, followed by $\mathcal{M}_{\rm 4pA}$ ($\Delta\ln\mathcal{Z}=-8$ against the 3-planet model), $\mathcal{M}_{\rm 4pB}$ ($\Delta\ln\mathcal{Z}=-17$), and $\mathcal{M}_{\rm 5p}$ ($\Delta\ln\mathcal{Z}=-21$).

%--------------------------
\subsection{Interior model for the three rocky worlds K2-32\,e, K2-233\,b \& K2-233\,c}

For the three planets with bulk densities compatible with rocky compositions, we can investigate their interior composition through different approaches. 

At first, we can estimate the core mass fraction (CMF) of these planets by using the analytic expressions in \cite{zeng19}. We find CMF$ = 71\pm65$\% for K2-32\,e, suggesting a scarce gaseous envelope and mantle for this dense rocky world, although still with large uncertainties given the precision obtained on its mass. For K2-233\,b we obtain CMF$ \sim$ 28\%, compatible with an Earth-like internal structure (CMF$_{\oplus}=33$\%) but also {with} large uncertainties, mainly due to the uncertainty in the planet mass. For K2-233\,c, we obtain CMF values outside of the validity range of the formula (CMF$\sim$90\%). However, given its location in the Mass-Radius diagram, a rocky composition is also very likely. Additional radial velocity monitoring, especially with ESPRESSO \citep{espresso}, will increase the significance on the {planet mass detections} and shed more light on their internal structure. 
In particular, in the case of K2-233, spectropolarimetric measurements with, e.g. the SPIRou spectrograph, {would help} to overcome the effects of stellar magnetic activity in such a young star \citep{hebrard14}. 

\begin{figure*}
\centering
\includegraphics[width=1\textwidth]{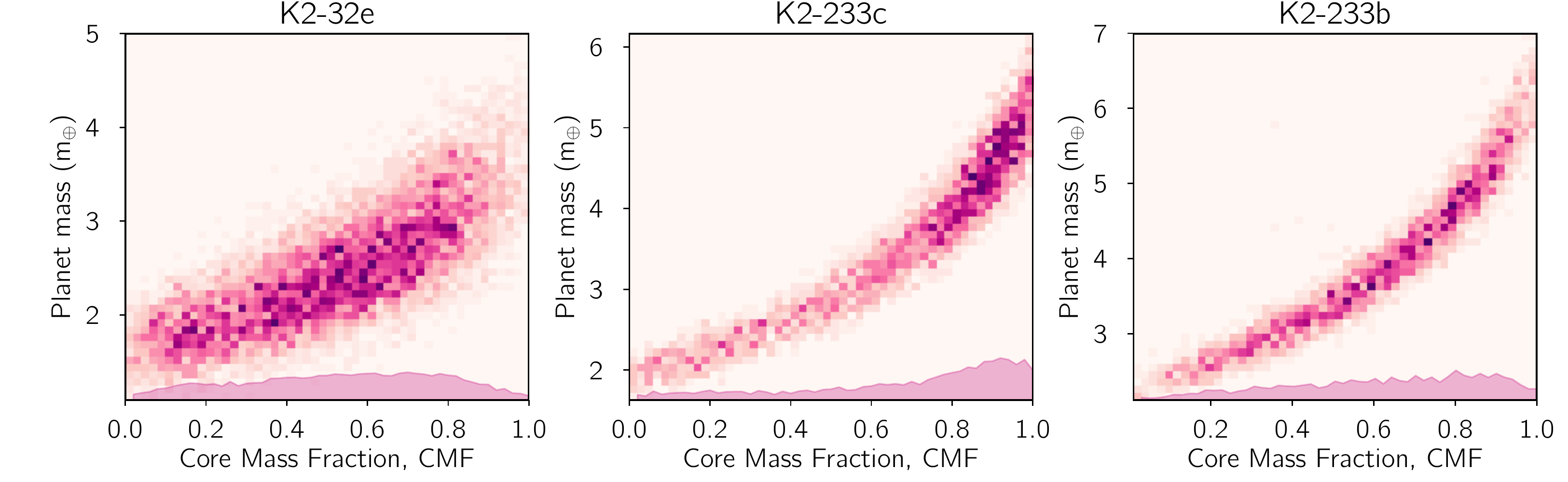} %/Users/lillo_box/00_projects/11__K2-32_K2-233/CMFplots/
\caption{{Sampled 2D marginal posterior distribution for the core mass function of K2-32\,e (left panel), K2-233\,c (middle panel), and K2-233\,b (right panel). Colour-code displays the normalised probability density function (PDF). At the bottom of each panel we show the one-dimensional PDF for the core mass fraction.}}
\label{fig:cmf}
\end{figure*}

\cite{dorn15} proposed that Mg/Si and Fe/Si mineralogical ratios can be used as probes to constrain the internal structure of terrestrial planets. Their theoretical models were successfully tested on three terrestrial planets by \cite{santos15} and used to explore the possible compositions of planet building blocks and planets orbiting stars belonging to different Galactic populations \citep{santos17}. Using the chemical abundances of K2-32 and K2-233 listed in Table~\ref{tab:specparams}, and assuming that C and O abundances for the two stars scale with metallicity, we applied the stoichiometric model of \cite{santos17} to determine the iron-mass fraction (which can be translated into core-mass fraction) of the planet building blocks in the {protoplanetary disks} of these stars. We note that solar-metallicity stars in the solar neighbourhood show solar C and O abundances \citep{bertran15,suarez-andres17}, {supporting our assumption of} C and O abundances. More importantly, the iron-mass fraction is not sensitive to C and O abundances \citep{santos17}. Our model suggests very similar iron mass fractions for {the planet building blocks in the two systems}: 32.8 $\pm$ 2.2\% and 32.8 $\pm$ 3.2 \% for K2-32 and K2-233, respectively. Note, that this model predicts an iron-mass fraction of 33.2\% for the solar system planet building blocks \citep{santos15}.

Finally, assuming an Earth-like albedo ($\sim$0.7), the equilibrium temperatures of K2-233\,b, K2-233\,c, and K2-32\,e are estimated to be $\sim$ {1169, 820}, and  1066~K, respectively. Therefore, if present, a hydrosphere would consist of a supercritical water phase surrounded with a hot and dense steam atmosphere \citep{mousis20}. However, using mass-radius relationships for planets harbouring water at high temperature, in a high pressure supercritical state as derived by \cite{mousis20}, we conclude that none of the three planets could harbour a significant hydrosphere. We thus assumed completely dry compositions, with a water mass fraction of 0, leaving the core mass fraction as the only compositional free parameter in our planet internal structure model \citep{brugger17}. The central values of the masses and radii of K2-233\,b, K2-233\,c, and K2-32\,e are consistent with a CMF of {CMF of 0.45, 0.94}, and 0.37, respectively. This means that K2-233b and K2-32e are compatible with an Earth-like composition, whereas K2-233c's composition is more Mercury-like (which resembles the case of K2-229\,b, see \citealt{santerne18}). We performed Monte-Carlo simulations to constrain the CMF range \citep{Dorn17}. We selected those CMF values whose input mass and output radius in our planet interior model are within the 1-$\sigma$ uncertainties of their measured values. The large uncertainties on the masses of these planets prevent a precise determination of the CMF which can vary between {0.29--0.82 for K2-233b,  between 0.22--0.92 for K2-32e, and between 0.78--1.0 for K2-233c (see probability density functions in Fig.~\ref{fig:cmf})}. Assessing more precise compositions and internal structure of these planets requires additional radial velocity measurements in order to obtain more accurate determinations of their masses. {Indeed, to reduce the current CMF errors by a factor of 2, it would be necessary to determine the mass with an uncertainty of $\sim$10\%. This would imply a significant effort for these three planets unaffordable with an instrument like HARPS (several thousands of measurements needed) and very costly for an ESPRESSO-like instrument ($>400$ measurements required).}

%________________________________________________________________
\section{Conclusions}
\label{sec:conclusions}

In this paper we measure the masses of seven planets in two planetary systems detected by the K2 mission (K2-32 and K2-233) through intense radial velocity monitoring with HARPS. We estimate for the first time the masses of K2-32\,e, K2-32\,c and constrain for the first time the masses of the three planets in the K2-233 system. We also provide improved measurements of {the masses of  K2-32\,b and K2-32\,d} {through a joint transit and radial velocity analysis with \pastis{} (see Tables~\ref{tab:k32sum} and \ref{tab:k233sum}).}

In K2-32 we find a wide diversity of properties among the four planets in the system with an architecture that resembles a scaled-down version of the Solar System, hosting an Earth-like planet in the inner region followed by a gas giant and two Neptune-like planets in longer period orbits. {Systems like K2-32} are very useful for dynamical studies to understand how different environments producing the same type of planets can build different architectures. The monitoring of the K2-233 system has allowed us for the first time to reach the rocky regime of planets around a young star ($<$1 Gyr). Our data {constrain the planets to be rocky in composition at high confidence}, although additional {measurements are} needed to increase the precision on their masses. 

Of special interest are the rocky and sub-Neptune planets of both systems, as they are transiting with {different orbital periods} around relatively bright stars. This positions them as {good candidates to be considered} for transmission spectroscopy studies, an unexplored parameter space that will soon be accessible with the \textit{James Webb} Space Telescope {\citep[see, e.g.,][]{malik19, mansfield19, koll19}.} 
 
\begin{acknowledgements}
This research has been funded by the Spanish State Research Agency (AEI) Projects No.ESP2017-87676-C5-1-R and No. MDM-2017-0737 Unidad de Excelencia "Mar\'ia de Maeztu"- Centro de Astrobiolog\'ia (INTA-CSIC). DJA acknowledges support from the STFC via an Ernest Rutherford Fellowship (ST/R00384X/1). This work was supported by FCT - Funda\c{c}\~ao para a Ci\^encia e a Tecnologia through national funds and by FEDER through COMPETE2020 - Programa Operacional Competitividade e Internacionaliza\c{c}\~ao by these grants: UID/FIS/04434/2019; UIDB/04434/2020; UIDP/04434/2020; PTDC/FIS-AST/32113/2017 \& POCI-01-0145-FEDER-032113; PTDC/FIS-AST/28953/2017 \& POCI-01-0145-FEDER-028953. SH acknowledge support by the fellowships PD/BD/128119/2016 funded by FCT (Portugal). SCCB, SGS, VA and ODSD acknowledge support from  Funda\c{c}\~ao para a Ci\^encia e a Tecnologia (FCT) through Investigador FCT contracts IF/01312/2014/CP1215/CT0004,  IF/00028/2014/CP1215/CT0002, IF/00650/2015/CP1273/CT0001 and DL 57/2016/CP1364/CT0004 respectively. 
\end{acknowledgements}

%-------------------------------------------------------------------
\bibliographystyle{aa} % style aa.bst
\bibliography{../biblio2} % your references Yourfile.bib

\appendix

\section{Tables}

\onecolumn

\input{Table_K2-32_LC__ktwo205071984c02_lpd_LC.tex}
\input{Table_K2-233_LC__ktwo249622103c15_lpd_LC.tex}

\input{Table_K2-32_RV.tex}
\input{Table_K2-233_RV.tex}

\newpage
\input{Table_K2-32_PASTIS_results.tex}

\newpage
\input{Table_K2-233_PASTIS_results.tex}

\end{document}

%% file: Table_SpectroscopicParams.tex
\begin{table}[]
\centering
\caption{Stellar parameters derived from the spectroscopic analysis (see Sect.~\ref{sec:stellar_params}).}
\label{tab:specparams}
\begin{tabular}{lcc}

\hline 
\textit{Stellar parameters} & \textbf{K2-32} &  \textbf{K2-233} \\ 
\hline 
&& \\
Eff. temp. \teff\ [K]                & $5273\pm25$          	& $4796\pm66$    \\ 
Surf. grav. \logg\ [cgs]                    & $4.359\pm0.048$      	& $4.53\pm0.22$   \\ 
Turb. vel. v$_t$ [km/s]                   & $0.617\pm0.057$ & $0.699\pm0.157$ \\
&& \\ 
\textit{Element abundances} & & \\
Fe/H  [dex]                      & $-0.055\pm0.017$                 		& $-0.082\pm0.028$    \\ 
Na  \Romannum{1} [dex]                      & $-0.012\pm0.108$                 		& $-0.165\pm0.084$    \\ 
Mg  \Romannum{1} [dex]                      & $-0.04\pm0.07$                 			& $-0.195\pm0.07$    \\ 
Al  \Romannum{1} [dex]                      & $0.104\pm0.092$                 		& $-0.036\pm0.068$    \\ 
Si  \Romannum{1} [dex]                      & $-0.053\pm0.041$                 		& $0.004\pm0.08$    \\ 
Ca  \Romannum{1} [dex]                      & $-0.045\pm0.048$                 		& $0.059\pm0.128$    \\ 
Ti  \Romannum{1} [dex]                      & $0.029\pm0.047$                 		& $-0.031\pm0.106$    \\ 
Cr  \Romannum{1} [dex]                      & $-0.021\pm0.047$                 		& $-0.068\pm0.102$    \\ 
Ni  \Romannum{1} [dex]                      & $-0.054\pm0.03$                 		& $-0.12\pm0.064$    \\ 
  &  &   \\ 
\hline 
 
 \end{tabular}
\end{table}

%% file: Table_Phot_mags.tex
  \begin{table}
  \begin{threeparttable}
    \caption{\label{tab:stellarparam1}Photometric magnitudes of K2-32 and K2-233.}
    \small
     \begin{tabular}{lcc}
        \toprule
         & K2-32 & K2-233 \\
        \midrule
        \hline
		\\
		Kepler Kp & $12.005^a$ & $10.422^a$\\ 
%		\hline
		Johnson B & $13.267\pm0.007^c$ & $11.664\pm0.027^c$\\
%		\hline
		Johnson V & $12.314\pm0.0043^c$ & $10.726\pm0.019^c$\\
%		\hline
		Sloan g$^{\prime}$ & $12.764\pm0.017^c$ & $11.173\pm0.043^c$\\
%		\hline
		Sloan r$^{\prime}$ & $12.005\pm0.026^c$ & $10.352\pm0.021^c$\\
%		\hline
		Sloan i$^{\prime}$ & $11.632\pm0.031^c$ & $10.068\pm0.019^c$\\
%		\hline
		2-MASS J & $10.404\pm0.024^b$ & $8.968\pm0.020^b$\\
%		\hline
		2-MASS H & $9.993\pm0.025^b$ & $8.501\pm0.026^b$\\
%		\hline
		2-MASS Ks & $9.821\pm0.019^b$ & $8.375\pm0.023^b$\\
%		\hline
		WISE W1 & $9.754\pm0.022^d$ & $8.318\pm0.023^d$\\
%		\hline
		WISE W2 & $9.787\pm0.020^d$ & $8.381\pm0.021^d$\\
%		\hline
		WISE W3 & $9.887\pm0.057^d$ & $9.332\pm0.025^d$\\
        \bottomrule
     \end{tabular}
    \begin{tablenotes}
      \small
      \item a. \url{https://exofop.ipac.caltech.edu/k2/}
      \item b. Two Micron All Sky Survey (2MASS)
      \item c. AAVSO Photometric All-Sky Survey (APASS)
      \item d. AllWISE
    \end{tablenotes}
  \end{threeparttable}
\end{table}

%% file: Table_Stellar_parameters_summary.tex
\begin{table}[]
\centering
\caption{Stellar parameters of K2-32 and K2-233 obtained from \texttt{PASTIS}.}
\label{tab:stellarpars}
\begin{tabular}{lcc}

\hline 
 \textit{Stellar parameters} & \textbf{K2-32} &  \textbf{K2-233} \\ 
\hline 
&& \\
Eff. temperature \teff\ [K]                		& $5271^{_{+39}}_{^{-35}}$          	& $5033^{_{+41}}_{^{-35}}$    \\ 
Surface gravity \logg\ [cgs]                    & $4.49^{_{+0.03}}_{^{-0.03}}$      	& $4.64\pm0.01$     \\ 
Iron abundance \met\ [dex]                      & $-0.06\pm0.03$                 		& $-0.09\pm0.03$   \\ 
Distance to Earth $D$ [pc]                      & $162.5^{_{+3.6}}_{^{-4.0}}$       	& $67.4\pm0.3$     \\ 
Extinction $E(B-V)$ [mag]                       & $0.162\pm0.019$                		& $0.031^{_{+0.021}}_{^{-0.018}}$   \\ 
Systemic RV $\gamma$ [\kms]                     & $-2.1983^{_{+0.0004}}_{^{-0.0005}}$ 	& $-9.6483^{_{+0.0055}}_{^{-0.0056}}$   \\ 
Stellar density $\rho_{\star}/\rho_{\astrosun}$ & $1.3^{_{+0.13}}_{^{-0.11}}$           & $2.25^{_{+0.04}}_{^{-0.03}}$    \\ 
Stellar mass M$_{\star}$\ [\Msun]               & $0.83\pm0.02$                  		& $0.79\pm0.01$     \\ 
Stellar radius R$_{\star}$\ [\Rsun]             & $0.86\pm0.02$                   		& $0.71\pm0.01$     \\ 
%Stellar age $\tau$\ [Gyr]                       & $11.2^{_{+2.7}}_{^{-2.9}}$         	& $0.6\pm0.3$     \\ 
Limb-darkening $u_{a}$                          & $0.496\pm0.009$                		& $0.551^{_{+0.010}}_{^{-0.011}}$   \\ 
Limb-darkening $u_{b}$                          & $0.204\pm0.006$                		& $0.165^{_{+0.008}}_{^{-0.007}}$    \\                    
  &  &   \\ 
\hline 
 
 \end{tabular}
\tablefoot{Stellar parameters given here were derived from \texttt{PASTIS} and are not from the spectral analysis. We assumed $\Rsun = 695\:508$ km, $\Msun = 1.98842 \times 10^{30}$ kg, $\Rearth = 6\:378\:137$ m }

\end{table}

%% file: Table_K2-32_summary_table.tex
\begin{table*}[]
\centering
\caption{System parameters of K2-32 obtained from the joint light curve and radial velocity analysis with \texttt{PASTIS}.}
\label{tab:k32sum}
\begin{tabular}{lcccccc}
 
 \hline 
{Planet parameters} &  K2-32\,\textbf{e} & K2-32\,\textbf{b} & K2-32\,\textbf{c} & K2-32\,\textbf{d} \\ 
\hline 
\\
Orbital Period $P$ [d] & $4.34934\pm0.00039$ & $8.99200\pm0.00008$ & $20.66093^{_{+0.00080}}_{^{-0.00079}}$ & $31.71701^{_{+0.00101}}_{^{-0.00096}}$ \\ 
$T_{0}$ [BJD - 2400000] & $56898.87863^{_{+0.00392}}_{^{-0.00412}}$ & $56900.92676\pm0.00031$ & $56899.42344^{_{+0.00167}}_{^{-0.00176}}$ & $56903.78707^{_{+0.00124}}_{^{-0.00123}}$ \\ 
RV semi-amplitude $K$ [\ms] & $0.92^{_{+0.57}}_{^{-0.51}}$ & $5.22\pm0.60$ & $2.15^{_{+0.64}}_{^{-0.65}}$ & $1.53\pm0.58$ \\ 
Orbital inclination $i$ [$^{\circ}$] & $89.0\pm0.7$ & $89.0^{_{+0.5}}_{^{-0.3}}$ & $89.4^{_{+0.3}}_{^{-0.2}}$ & $89.4\pm0.1$ \\ 
Planet-to-star radius ratio $k$ & $0.01294^{_{+0.00042}}_{^{-0.00040}}$ & $0.05642^{_{+0.00071}}_{^{-0.00066}}$ & $0.03342^{_{+0.00061}}_{^{-0.00049}}$ & $0.03704^{_{+0.00062}}_{^{-0.00054}}$ \\ 
Orbital eccentricity $e$ & $0.043^{_{+0.048}}_{^{-0.030}}$ & $0.030^{_{+0.032}}_{^{-0.021}}$ & $0.049^{_{+0.046}}_{^{-0.035}}$ & $0.050^{_{+0.053}}_{^{-0.035}}$ \\ 
Argument of periastron $\omega$ [$^{\circ}$] & $205^{_{+106}}_{^{-134}}$ & $220^{_{+77}}_{^{-161}}$ & $168^{_{+117}}_{^{-103}}$ & $185^{_{+105}}_{^{-104}}$ \\ 
System scale $a/R_{\star}$ & $12.24^{_{+0.40}}_{^{-0.36}}$ & $19.86^{_{+0.65}}_{^{-0.59}}$ & $34.58^{_{+1.13}}_{^{-1.03}}$ & $46.02^{_{+1.50}}_{^{-1.37}}$ \\ 
Impact parameter $b$ & $0.22\pm0.16$ & $0.34^{_{+0.10}}_{^{-0.17}}$ & $0.37^{_{+0.11}}_{^{-0.17}}$ & $0.46^{_{+0.08}}_{^{-0.10}}$ \\ 
Transit duration T$_{14}$ [h] & $2.671^{_{+0.096}}_{^{-0.095}}$ & $3.488^{_{+0.022}}_{^{-0.019}}$ & $4.350^{_{+0.046}}_{^{-0.047}}$ & $4.888^{_{+0.051}}_{^{-0.048}}$ \\ 
Semi-major axis $a$ [AU] & $0.04899^{_{+0.00041}}_{^{-0.00038}}$ & $0.07950^{_{+0.00066}}_{^{-0.00062}}$ & $0.13843^{_{+0.00115}}_{^{-0.00108}}$ & $0.18422^{_{+0.00152}}_{^{-0.00144}}$ \\ 
Planet mass M [\Mearth] & $2.1^{_{+1.3}}_{^{-1.1}}$ & $15.0^{_{+1.8}}_{^{-1.7}}$ & $8.1\pm2.4$ & $6.7\pm2.5$ \\ 
Planet radius R [\Rearth] & $1.212^{_{+0.052}}_{^{-0.046}}$ & $5.299\pm0.191$ & $3.134^{_{+0.123}}_{^{-0.102}}$ & $3.484^{_{+0.112}}_{^{-0.129}}$ \\ 
Planet bulk density $\rho$ [\gcm3] & $6.30^{_{+4.13}}_{^{-3.49}}$ & $0.55^{_{+0.10}}_{^{-0.08}}$ & $1.43^{_{+0.48}}_{^{-0.45}}$ & $0.87^{_{+0.35}}_{^{-0.34}}$ \\ 
Mean Eq. temperature $T_{eq}$ [K] & $1066^{_{+16}}_{^{-18}}$ & $837^{_{+13}}_{^{-14}}$ & $634^{_{+9}}_{^{-11}}$ & $550^{_{+8}}_{^{-9}}$ \\ 
Day-side temperature $T_{eq, d}$ [K] & $1091^{_{+30}}_{^{-23}}$ & $850^{_{+21}}_{^{-18}}$ & $650^{_{+19}}_{^{-15}}$ & $564^{_{+20}}_{^{-14}}$ \\ 
&&&&&& \\
 
\hline

 \end{tabular}
\tablefoot{Stellar parameters from Table~\ref{tab:stellarpars} (derived from \texttt{PASTIS} and not from the spectral analysis). We assumed $\Rsun = 695\:508$ km, $\Msun = 1.98842 \times 10^{30}$ kg, $\Rearth = 6\:378\:137$ m, $\Mearth = 5.9736 \times 10^{24}$ kg and 1 AU $= 149\:597\:870.7$ km. The temperatures were derived assuming a zero albedo. The day-side temperature was computed assuming tidally synchronized rotation.}

\end{table*}

%% file: Table_K2-233_summary_table.tex
\begin{table*}[]
%\centering
%\setlength{\extrarowheight}{5pt}
\caption{\label{tab:k233sum}System parameters of K2-233 obtained from   the joint light curve and radial velocity analysis with \texttt{PASTIS}.}
\begin{tabular}{lcccccc}
 
 \hline 
{Planet parameters} & K2-233\,\textbf{b} & K2-233\,\textbf{c} & K2-233\,\textbf{d} \\ 
\hline 
\\
Orbital Period $P$ [d] & $2.46750\pm0.00004$ & $7.06005^{_{+0.00024}}_{^{-0.00027}}$ & $24.36450\pm0.00068$ \\ 
$T_{0}$ [BJD - 2400000] & $57991.69112^{_{+0.00083}}_{^{-0.00084}}$ & $57586.87653^{_{+0.01769}}_{^{-0.01567}}$ & $58005.58237^{_{+0.00092}}_{^{-0.00088}}$ \\ 
RV semi-amplitude $K$ [\ms] & $1.86^{_{+1.65}}_{^{-1.22}} (<6.27)$& $1.99^{_{+1.26}}_{^{-1.14}} (<5.02)$ & $2.13^{_{+1.36}}_{^{-1.20}} (<5.45)$ \\ 
Orbital inclination $i$ [$^{\circ}$] & $89.3^{_{+0.5}}_{^{-0.7}}$ & $89.6^{_{+0.3}}_{^{-0.4}}$ & $89.6^{_{+0.2}}_{^{-0.1}}$ \\ 
Planet-to-star radius ratio $k$ & $0.01743^{_{+0.00019}}_{^{-0.00016}}$ & $0.01663^{_{+0.00026}}_{^{-0.00024}}$ & $0.03062^{_{+0.00048}}_{^{-0.00040}}$ \\ 
Orbital eccentricity $e$ & $0.079^{_{+0.048}}_{^{-0.029}}$ & $0.070^{_{+0.045}}_{^{-0.033}} (<0.2002)$ & $0.062^{_{+0.043}}_{^{-0.042}} (<0.1999)$ \\ 
Argument of periastron $\omega$ [$^{\circ}$] & $248^{_{+53}}_{^{-35}}$ & $261^{_{+52}}_{^{-44}}$ & $139^{_{+100}}_{^{-82}}$ \\ 
System scale $a/R_{\star}$ & $10.07\pm0.05$ & $20.30^{_{+0.11}}_{^{-0.10}}$ & $46.36\pm0.24$ \\ 
Impact parameter $b$ & $0.13^{_{+0.13}}_{^{-0.09}}$ & $0.13^{_{+0.14}}_{^{-0.09}}$ & $0.32^{_{+0.11}}_{^{-0.20}}$ \\ 
Transit duration T$_{14}$ [h] & $1.988\pm0.026$ & $2.796^{_{+0.057}}_{^{-0.053}}$ & $3.838\pm0.033$ \\ 
Semi-major axis $a$ [AU] & $0.03308^{_{+0.00010}}_{^{-0.00009}}$ & $0.06666^{_{+0.00020}}_{^{-0.00018}}$ & $0.15224^{_{+0.00046}}_{^{-0.00042}}$ \\ 
Planet mass M [\Mearth] & $3.3^{_{+3.0}}_{^{-2.2}} (<11.26)$ & $5.1^{_{+3.2}}_{^{-2.9}} (<12.81)$ & $8.3^{_{+5.2}}_{^{-4.7}} (<21.14)$ \\ 
Planet radius R [\Rearth] & $1.343^{_{+0.018}}_{^{-0.016}}$ & $1.281^{_{+0.022}}_{^{-0.021}}$ & $2.358^{_{+0.043}}_{^{-0.036}}$ \\ 
Planet bulk density $\rho$ [\gcm3] & $7.59^{_{+6.74}}_{^{-5.00}} (<26.26)$ & $13.30^{_{+8.49}}_{^{-7.60}} (<33.91)$ & $3.44^{_{+2.20}}_{^{-1.94}} (<8.85)$ \\ 
Mean Eq. temperature $T_{eq}$ [K] & $1121^{_{+10}}_{^{-9}}$ & $790^{_{+7}}_{^{-6}}$ & $523^{_{+5}}_{^{-4}}$ \\ 
Day-side temperature $T_{eq, d}$ [K] & $1169^{_{+32}}_{^{-20}}$ & $820^{_{+21}}_{^{-15}}$ & $540^{_{+15}}_{^{-12}}$ \\ 
&&& \\

\hline

 \end{tabular}
\tablefoot{Stellar parameters from Table~\ref{tab:stellarpars} (derived from \texttt{PASTIS} and not from the spectral analysis). For the mass, we report the three sigma credible intervals in parenthesis. We assumed $\Rsun = 695\:508$ km, $\Msun = 1.98842 \times 10^{30}$ kg, $\Rearth = 6\:378\:137$ m, $\Mearth = 5.9736 \times 10^{24}$ kg and 1 AU $= 149\:597\:870.7$ km. The temperatures were derived assuming a zero albedo. The day-side temperature was computed assuming tidally synchronized rotation.}

\end{table*}

%% file: Table_K2-32_LC__ktwo205071984c02_lpd_LC.tex
\begin{table}[]
\centering
\caption{K2 photometry for K2-32. Only the first 10 measurements are shown. The complete table can be found in the online version of the article and in the CDS.}
\label{tab:k2-32_lc}
\begin{tabular}{lll}

\hline 
JD-2400000 (days) & Rel. Flux &  $\sigma_{\rm F}$ \\ 
\hline 

        56896.0020686457937700  &  1.0100155528817736  & 0.0000578899791713 \\
        56896.0225009107452934  &  1.0102177384327031  & 0.0000578975736780 \\
        56896.0429331747582182  &  1.0099493768616579  & 0.0000578915433817 \\
        56896.0633654387784190  &  1.0099700949243571  & 0.0000578940184963 \\
        56896.0837977023329586  &  1.0099545127464908  & 0.0000578957638977 \\
        56896.1042300654153223  &  1.0099669385815275  & 0.0000578981458503 \\
        56896.1246623275728780  &  1.0098439191841280  & 0.0000578955898924 \\
        56896.1450944897296722  &  1.0097765358677386  & 0.0000578962775403 \\
        56896.1655268514150521  &  1.0100921635003299  & 0.0000579076950933 \\
        56896.1859591121756239  &  1.0099820017896914  & 0.0000579078666161 \\
        56896.2063912729354342  &  1.0100788691335243  & 0.0000579146553820 \\

... & ... & ...  \\
\hline 
 
 \end{tabular}
\end{table}

%% file: Table_K2-233_LC__ktwo249622103c15_lpd_LC.tex
\begin{table}[]
\centering
\caption{K2 photometry for K2-233. Only the first 10 measurements are shown. The complete table can be found in the online version of the article and in the CDS.}
\label{tab:k2-233_lc}
\begin{tabular}{lll}

\hline 
JD-2400000 (days) & Rel. Flux &  $\sigma_{\rm F}$  \\ 
\hline 
        57989.4448759220031206   &    0.9924592443158369    &   0.000026376146163 \\       
        57989.4653083558732760   &    0.9926550253234170    &   0.000026381437803 \\       
        57989.4857406892770086   &    0.9928540924140880    &   0.000026385065025 \\       
        57989.5061729222070426   &    0.9929645697508946    &   0.000026387697919 \\       
        57989.5266053546802141   &    0.9929870028323399    &   0.000026388964194 \\       
        57989.5470376862213016   &    0.9929040218922018    &   0.000026387612766 \\       
        57989.5674701177558745   &    0.9929559809328470    &   0.000026389107223 \\       
        57989.5879023488305393   &    0.9929432366928289    &   0.000026389497357 \\       
        57989.6083346789746429   &    0.9928129961406429    &   0.000026387438345 \\       
        57989.6287671091122320   &    0.9930717947025098    &   0.000026389007022 \\       
... & ... & ...  \\
\hline 
 
 \end{tabular}
\end{table}

%% file: Table_K2-32_RV.tex
\begin{table}[]
\centering
\caption{Radial velocities for K2-32 used for this study. Only the first 10 measurements are shown. The complete table can be found in the online version of the article and in the CDS.}
\label{tab:k2-32_rv}
\begin{tabular}{llll}

\hline 
JD-2400000 (days) & RV (km/s) &  $\sigma_{\rm RV}$ (km/s) & Instrument \\ 
\hline 
57185.606905 &	-1.86757 &	0.00292	&     HARPS \\
57185.677182 &	-1.86511 &	0.00264	&     HARPS \\
57185.819738 &	-1.86373 &	0.00313	&     HARPS \\
57185.849749 &	-1.86628 &	0.00510	&     HARPS \\
57186.599740 &	-1.87200 &	0.00388	&     HARPS \\
57186.669808 &	-1.86611 &	0.00328	&     HARPS \\
57186.790154 &	-1.87182 &	0.00325	&     HARPS \\
57186.823625 &	-1.86549 &	0.00363	&     HARPS \\
57187.616519 &	-1.86562 &	0.00235	&     HARPS \\
57187.674585 &	-1.86930 &	0.00207	&     HARPS \\
57187.743821 &	-1.87361 &	0.00280	&     HARPS \\
... & ... & ... &... \\

\hline 
 
 \end{tabular}
\end{table}

%% file: Table_K2-233_RV.tex
\begin{table}[]
\centering
\caption{Radial velocities for K2-233 used for this study. Only the first 10 measurements are shown. The complete table can be found in the online version of the article and in the CDS.}
\label{tab:k2-233_rv}
\begin{tabular}{llll}

\hline 
JD-2400000 (days) & RV (km/s) &  $\sigma_{\rm RV}$ (km/s) & Instrument \\ 
\hline 
58257.575544860 &	-9.62195 &	0.00190 &	HARPS \\
58257.636215759 &	-9.62438 &	0.00202 &	HARPS \\
58257.754999690 &	-9.62594 &	0.00200 &	HARPS \\
58258.564438059 &	-9.65918 &	0.00185 &	HARPS \\
58258.658789129 &	-9.66632 &	0.00218 &	HARPS \\
58259.617947040 &	-9.67847 &	0.00183 &	HARPS \\
58259.736962050 &	-9.67739 &	0.00182 &	HARPS \\
58260.580031630 &	-9.65730 &	0.00161 &	HARPS \\
58260.831405989 &	-9.65559 &	0.00191 &	HARPS \\
58261.660944140 &	-9.65916 &	0.00179 &	HARPS \\
58261.822029939 &	-9.65921 &	0.00177 &	HARPS \\
... & ... & ... &... \\
\hline 
 
 \end{tabular}
\end{table}

%% file: Table_K2-32_PASTIS_results.tex
\begin{landscape}
\begin{longtable}{lccc}
\caption{\label{tab:K32_MCMC} List of parameters used in the analysis of K2-32 planets. The respective priors are provided together with the posteriors for the Dartmouth and PARSEC stellar evolution tracks. The posterior values represent the median and 68.3\% credible interval. Derived values that might be useful for follow-up work are also reported.}\\
\hline
Parameter & Prior & \multicolumn{2}{c}{Posterior}\\
 &  & Dartmouth & PARSEC\\
 &  & (adopted) & \\
\hline
\endfirsthead
\multicolumn{4}{l}{{\bfseries \tablename\ \thetable{} -- continued from previous page}} \\
\hline
Parameter & Prior & \multicolumn{2}{c}{Posterior}\\
 &  & Dartmouth & PARSEC\\
\hline
\endhead
\multicolumn{4}{l}{{Continued on next page}} \\ 
\hline
\endfoot
\hline
\multicolumn{4}{l}{Notes:}\\
\multicolumn{4}{l}{$\bullet$ $\mathcal{N}(\mu,\sigma^{2})$: Normal distribution with mean $\mu$ and width $\sigma^{2}$}\\
\multicolumn{4}{l}{$\bullet$ $\mathcal{U}(a,b)$: Uniform distribution between $a$ and $b$}\\
\multicolumn{4}{l}{$\bullet$ $\mathcal{S}(a,b)$: Sine distribution between $a$ and $b$, corresponding to $\mathcal{U}(\sin{a},\sin{b})$}\\
%\multicolumn{4}{l}{$\bullet$ $\mathcal{N}_{\mathcal{U}}(\mu,\sigma^{2},a,b)$: Normal distribution with mean $\mu$ and width $\sigma^{2}$ multiplied with a uniform distribution between $a$ and $b$}\\
%\multicolumn{4}{l}{$\bullet$ $\mathcal{S}(a,b)$: Sine distribution between $a$ and $b$}\\
\multicolumn{4}{l}{$\bullet$ $\mathcal{T}(\mu,\sigma^{2},a,b)$: Truncated normal distribution with mean $\mu$ and width $\sigma^{2}$, between $a$ and $b$}\\
\endlastfoot
\\
\multicolumn{4}{l}{\it Stellar Parameters}\\
\\
Effective temperature \teff\ [K] & $\mathcal{N}(5276, 37)$ & $5271.2^{_{+39.1}}_{^{-35.2}}$ & $5269.4^{_{+34.8}}_{^{-31.3}}$ \\
Surface gravity \logg\ [cgs] & $\mathcal{N}(4.38,0.06)$ & $4.49^{_{+0.03}}_{^{-0.03}}$ & $4.47^{_{+0.03}}_{^{-0.03}}$ \\
Iron abundance \met\ [dex] &  $\mathcal{N}(-0.05,0.03)$ & $-0.06\pm0.03$ & $-0.06\pm0.03$ \\
Distance to Earth $D$ [pc] &  $\mathcal{N}(157.118,4.97)$ & $162.5^{_{+3.6}}_{^{-4.0}}$ & $160.9^{_{+4.2}}_{^{-4.1}}$ \\
Interstellar extinction $E(B-V)$ [mag] &  $\mathcal{U}(0.0,1.0)$ & $0.162^{_{+0.019}}_{^{-0.019}}$ & $0.158^{_{+0.019}}_{^{-0.019}}$ \\
Systemic radial velocity $\gamma$ [\kms] & $\mathcal{U}(-5.0,0.0)$ & $-2.1983^{_{+0.0004}}_{^{-0.0005}}$ & $-2.1983^{_{+0.0004}}_{^{-0.0005}}$ \\
Linear limb-darkening coefficient $u_{a}$ & (derived) & $0.4964^{_{+0.0088}}_{^{-0.0093}}$ & $0.4966^{_{+0.0084}}_{^{-0.0082}}$ \\
Quadratic limb-darkening coefficient $u_{b}$ & (derived) & $0.2042^{_{+0.006}}_{^{-0.006}}$ & $0.204^{_{+0.0054}}_{^{-0.0056}}$ \\
Stellar density $\rho_{\star}/\rho_{\astrosun}$ & (derived) & $1.299^{_{+0.131}}_{^{-0.112}}$ & $1.270^{_{+0.114}}_{^{-0.099}}$ \\
Stellar mass M$_{\star}$\ [\Msun] & (derived) & $0.829^{_{+0.021}}_{^{-0.019}}$ & $0.78^{_{+0.018}}_{^{-0.018}}$ \\
Stellar radius R$_{\star}$\ [\Rsun] & (derived) & $0.861^{_{+0.022}}_{^{-0.023}}$ & $0.849^{_{+0.022}}_{^{-0.02}}$ \\
Stellar age $\tau$\ [Gyr] & (derived) & $11.2^{_{+2.7}}_{^{-2.9}}$ & $15.9^{_{+2.5}}_{^{-3.0}}$ \\
&&& \\
\hline
\\
\multicolumn{4}{l}{\it Planet e Parameters}\\
\\
Orbital Period $P_{e}$ [d] &  $\mathcal{N}(4.34882,0.07)$ & $4.34934^{_{+0.00039}}_{^{-0.00039}}$ & $4.34933^{_{+0.00039}}_{^{-0.00037}}$ \\
Transit epoch $T_{0,e}$ [BJD - 2450000] &  $\mathcal{N}(6898.886,0.1)$ & $6898.87863^{_{+0.00392}}_{^{-0.00412}}$ & $6898.87870^{_{+0.00358}}_{^{-0.00398}}$ \\
Radial velocity semi-amplitude $K_{e}$ [\ms] & $\mathcal{U}(0,100)$ & $0.92^{_{+0.57}}_{^{-0.51}}$ & $0.92^{_{+0.58}}_{^{-0.51}}$ \\
Orbital inclination $i_{e}$ [$^{\circ}$] & $\mathcal{S}(70.0,90.0)$ & $89.0\pm0.7$ & $88.9\pm0.7$ \\
Planet-to-star radius ratio $k_{e}$ & $\mathcal{U}(0.0,0.2)$ & $0.01294^{_{+0.00042}}_{^{-0.00040}}$ & $0.01299^{_{+0.00042}}_{^{-0.00041}}$ \\
Orbital eccentricity $e_{e}$ & $\mathcal{T}(0.0,0.083,0.0,1.0)$ & $0.043^{_{+0.048}}_{^{-0.030}}$ & $0.041^{_{+0.047}}_{^{-0.030}}$ \\
Argument of periastron $\omega_{e}$ [\degr] & $\mathcal{U}(0.0,360.0)$ & $205^{_{+106}}_{^{-134}}$ & $158^{_{+157}}_{^{-115}}$ \\
System scale $a_{e}/R_{\star}$ & (derived) & $12.2^{_{+0.4}}_{^{-0.4}}$ & $12.1^{_{+0.4}}_{^{-0.3}}$ \\
Impact parameter $b_{e}$ & (derived) & $0.220^{_{+0.164}}_{^{-0.156}}$ & $0.229^{_{+0.154}}_{^{-0.142}}$ \\
Transit duration T$_{14,e}$ [h] & (derived) & $2.67^{_{+0.10}}_{^{-0.10}}$ & $2.66^{_{+0.09}}_{^{-0.10}}$ \\
Semi-major axis $a_{e}$ [AU] & (derived) & $0.04899^{_{+0.00041}}_{^{-0.00038}}$ & $0.04799^{_{+0.00036}}_{^{-0.00037}}$ \\
Planet mass M$_{e}$ [\Mearth] & (derived) & $2.07^{_{+1.26}}_{^{-1.14}}$ & $1.99^{_{+1.25}}_{^{-1.10}}$ \\
Planet radius R$_{e}$ [\Rearth] & (derived) & $1.212^{_{+0.052}}_{^{-0.046}}$ & $1.202^{_{+0.052}}_{^{-0.047}}$ \\
Planet bulk density $\rho_{e}$ [\gcm3] & (derived) & $6.3^{_{+4.1}}_{^{-3.5}}$ & $6.2^{_{+4.1}}_{^{-3.5}}$ \\
&&& \\
\hline
\\
\multicolumn{4}{l}{\it Planet b Parameters}\\
\\
Orbital Period $P_{b}$ [d] &  $\mathcal{N}(8.991828,0.008)$ & $8.99200\pm0.00008$ & $8.99201\pm0.00008$ \\
Transit epoch $T_{0,b}$ [BJD - 2450000] &  $\mathcal{N}(6900.92713,0.03)$ & $6900.92676^{_{+0.00031}}_{^{-0.00031}}$ & $6900.92675^{_{+0.00032}}_{^{-0.00031}}$ \\
Radial velocity semi-amplitude $K_{b}$ [\ms] & $\mathcal{U}(0,100)$ & $5.22\pm0.60$ & $5.22\pm0.60$ \\
Orbital inclination $i_{b}$ [$^{\circ}$] & $\mathcal{S}(70.0,90.0)$ & $89.0^{_{+0.5}}_{^{-0.3}}$ & $89.0^{_{+0.3}}_{^{-0.2}}$ \\
Planet-to-star radius ratio $k_{b}$ & $\mathcal{U}(0.0,0.2)$ & $0.05642^{_{+0.00071}}_{^{-0.00066}}$ & $0.05642^{_{+0.00057}}_{^{-0.00051}}$ \\
Orbital eccentricity $e_{b}$ & $\mathcal{T}(0.0,0.083,0.0,1.0)$ & $0.030^{_{+0.032}}_{^{-0.021}}$ & $0.033^{_{+0.034}}_{^{-0.023}}$ \\
Argument of periastron $\omega_{b}$ [\degr] & $\mathcal{U}(0.0,360.0)$ & $220^{_{+77}}_{^{-161}}$ & $178^{_{+89}}_{^{-69}}$ \\
System scale $a_{b}/R_{\star}$ & (derived) & $19.9^{_{+0.6}}_{^{-0.6}}$ & $19.7^{_{+0.6}}_{^{-0.5}}$ \\
Impact parameter $b_{b}$ & (derived) & $0.338^{_{+0.101}}_{^{-0.167}}$ & $0.338^{_{+0.084}}_{^{-0.103}}$ \\
Transit duration T$_{14,b}$ [h] & (derived) & $3.49\pm0.02$ & $3.49\pm0.02$ \\
Semi-major axis $a_{b}$ [AU] & (derived) & $0.07950^{_{+0.00066}}_{^{-0.00062}}$ & $0.07789^{_{+0.00059}}_{^{-0.00060}}$ \\
Planet mass M$_{b}$ [\Mearth] & (derived) & $14.97^{_{+1.79}}_{^{-1.73}}$ & $14.36^{_{+1.69}}_{^{-1.61}}$ \\
Planet radius R$_{b}$ [\Rearth] & (derived) & $5.299^{_{+0.191}}_{^{-0.191}}$ & $5.223^{_{+0.176}}_{^{-0.153}}$ \\
Planet bulk density $\rho_{b}$ [\gcm3] & (derived) & $0.6\pm0.1$ & $0.6\pm0.1$ \\
&&& \\
\hline
\\
\multicolumn{4}{l}{\it Planet c Parameters}\\
\\
Orbital Period $P_{c}$ [d] &  $\mathcal{N}(20.66186,0.1)$ & $20.66093^{_{+0.00080}}_{^{-0.00079}}$ & $20.66089^{_{+0.00082}}_{^{-0.00080}}$ \\
Transit epoch $T_{0,c}$ [BJD - 2450000] &  $\mathcal{N}(6899.4227,0.1)$ & $6899.42344^{_{+0.00167}}_{^{-0.00176}}$ & $6899.42348^{_{+0.00167}}_{^{-0.00181}}$ \\
Radial velocity semi-amplitude $K_{c}$ [\ms] & $\mathcal{U}(0,100)$ & $2.15^{_{+0.64}}_{^{-0.65}}$ & $2.22^{_{+0.64}}_{^{-0.62}}$ \\
Orbital inclination $i_{c}$ [$^{\circ}$] & $\mathcal{S}(70.0,90.0)$ & $89.4^{_{+0.3}}_{^{-0.2}}$ & $89.3^{_{+0.2}}_{^{-0.1}}$ \\
Planet-to-star radius ratio $k_{c}$ & $\mathcal{U}(0.0,0.2)$ & $0.03342^{_{+0.00061}}_{^{-0.00049}}$ & $0.03362^{_{+0.00051}}_{^{-0.00048}}$ \\
Orbital eccentricity $e_{c}$ & $\mathcal{T}(0.0,0.083,0.0,1.0)$ & $0.049^{_{+0.046}}_{^{-0.035}}$ & $0.042^{_{+0.041}}_{^{-0.029}}$ \\
Argument of periastron $\omega_{c}$ [\degr] & $\mathcal{U}(0.0,360.0)$ & $168^{_{+117}}_{^{-103}}$ & $191^{_{+112}}_{^{-130}}$ \\
System scale $a_{c}/R_{\star}$ & (derived) & $34.6^{_{+1.1}}_{^{-1.0}}$ & $34.3^{_{+1.0}}_{^{-0.9}}$ \\
Impact parameter $b_{c}$ & (derived) & $0.370^{_{+0.112}}_{^{-0.168}}$ & $0.412^{_{+0.083}}_{^{-0.106}}$ \\
Transit duration T$_{14,c}$ [h] & (derived) & $4.35\pm0.05$ & $4.35\pm0.04$ \\
Semi-major axis $a_{c}$ [AU] & (derived) & $0.13843^{_{+0.00115}}_{^{-0.00108}}$ & $0.13562^{_{+0.00103}}_{^{-0.00105}}$ \\
Planet mass M$_{c}$ [\Mearth] & (derived) & $8.13^{_{+2.41}}_{^{-2.45}}$ & $8.05^{_{+2.31}}_{^{-2.25}}$ \\
Planet radius R$_{c}$ [\Rearth] & (derived) & $3.134^{_{+0.123}}_{^{-0.102}}$ & $3.110^{_{+0.113}}_{^{-0.097}}$ \\
Planet bulk density $\rho_{c}$ [\gcm3] & (derived) & $1.4^{_{+0.5}}_{^{-0.4}}$ & $1.5^{_{+0.5}}_{^{-0.4}}$ \\
&&& \\
\hline
\\
\multicolumn{4}{l}{\it Planet d Parameters}\\
\\
Orbital Period $P_{d}$ [d] &  $\mathcal{N}(31.7142,0.1)$ & $31.71701^{_{+0.00101}}_{^{-0.00096}}$ & $31.71697^{_{+0.00102}}_{^{-0.00093}}$ \\
Transit epoch $T_{0,d}$ [BJD - 2450000] &  $\mathcal{N}(6903.7913,0.1)$ & $6903.78707^{_{+0.00124}}_{^{-0.00123}}$ & $6903.78708^{_{+0.00120}}_{^{-0.00122}}$ \\
Radial velocity semi-amplitude $K_{d}$ [\ms] & $\mathcal{U}(0,100)$ & $1.53^{_{+0.58}}_{^{-0.58}}$ & $1.51^{_{+0.58}}_{^{-0.57}}$ \\
Orbital inclination $i_{d}$ [$^{\circ}$] & $\mathcal{S}(70.0,90.0)$ & $89.4\pm0.1$ & $89.4\pm0.1$ \\
Planet-to-star radius ratio $k_{d}$ & $\mathcal{U}(0.0,0.2)$ & $0.03704^{_{+0.00062}}_{^{-0.00054}}$ & $0.03718^{_{+0.00067}}_{^{-0.00058}}$ \\
Orbital eccentricity $e_{d}$ & $\mathcal{T}(0.0,0.083,0.0,1.0)$ & $0.050^{_{+0.053}}_{^{-0.035}}$ & $0.047^{_{+0.058}}_{^{-0.033}}$ \\
Argument of periastron $\omega_{d}$ [\degr] & $\mathcal{U}(0.0,360.0)$ & $185^{_{+105}}_{^{-104}}$ & $198^{_{+125}}_{^{-131}}$ \\
System scale $a_{d}/R_{\star}$ & (derived) & $46.0^{_{+1.5}}_{^{-1.4}}$ & $45.7^{_{+1.3}}_{^{-1.2}}$ \\
Impact parameter $b_{d}$ & (derived) & $0.457^{_{+0.083}}_{^{-0.100}}$ & $0.480^{_{+0.083}}_{^{-0.096}}$ \\
Transit duration T$_{14,d}$ [h] & (derived) & $4.89^{_{+0.05}}_{^{-0.05}}$ & $4.89^{_{+0.06}}_{^{-0.05}}$ \\
Semi-major axis $a_{d}$ [AU] & (derived) & $0.18422^{_{+0.00152}}_{^{-0.00144}}$ & $0.18048^{_{+0.00137}}_{^{-0.00140}}$ \\
Planet mass M$_{d}$ [\Mearth] & (derived) & $6.66^{_{+2.54}}_{^{-2.52}}$ & $6.30^{_{+2.41}}_{^{-2.42}}$ \\
Planet radius R$_{d}$ [\Rearth] & (derived) & $3.484^{_{+0.112}}_{^{-0.129}}$ & $3.443^{_{+0.142}}_{^{-0.117}}$ \\
Planet bulk density $\rho_{d}$ [\gcm3] & (derived) & $0.9\pm0.3$ & $0.8\pm0.3$ \\
&&& \\
\hline
&&& \\
\multicolumn{4}{l}{\it Instrument-related Parameters}\\
&&& \\
HARPS jitter $\sigma_{j,~\rm RV}$ [\ms] & $\mathcal{U}(0,100)$ & $4.03^{_{+0.43}}_{^{-0.39}}$ & $3.99^{_{+0.46}}_{^{-0.40}}$ \\
HIRES jitter $\sigma_{j,~\rm RV}$ [\ms] & $\mathcal{U}(0,100)$ & $5.65^{_{+0.97}}_{^{-0.77}}$ & $5.79^{_{+0.98}}_{^{-0.80}}$ \\
PFS jitter $\sigma_{j,~\rm RV}$ [\ms] & $\mathcal{U}(0,100)$ & $0.006.29^{_{+3.68}}_{^{-2.24}}$ & $6.3^{_{+3.83}}_{^{-2.26}}$ \\
HIRES offset [\kms] & $\mathcal{U}(-5.0,5.0)$ & $-2.1977^{_{+0.00117}}_{^{-0.00118}}$ & $-2.19761^{_{+0.00121}}_{^{-0.00121}}$ \\
PFS offset [\kms] & $\mathcal{U}(-5.0,5.0)$ & $-2.19098^{_{+0.00298}}_{^{-0.00282}}$ & $-2.19104^{_{+0.00293}}_{^{-0.00293}}$ \\
\textit{K2} contamination [\%] & $\mathcal{T}(0.0,0.005,0.0,1.0)$ & $0.003^{_{+0.004}}_{^{-0.002}}$ & $0.003^{_{+0.004}}_{^{-0.002}}$ \\
\textit{K2} jitter $\sigma_{j,~\textit{K2}}$ [ppm] & $\mathcal{U}(0.0, 10^5)$ & $63.4\pm1.7$ & $63.4\pm1.6$ \\
\textit{K2} out-of-transit flux & $\mathcal{U}(0.99,1.01)$ & $1.0000003^{_{+0.0000017}}_{^{-0.0000016}}$ & $1.0000003^{_{+0.0000016}}_{^{-0.0000016}}$ \\
SED jitter [mag]  & $\mathcal{U}(0.0,0.1)$ & $0.071^{_{+0.017}}_{^{-0.015}}$ & $0.07^{_{+0.018}}_{^{-0.014}}$ \\
&&& \\
\end{longtable}
\end{landscape}

%% file: Table_K2-233_PASTIS_results.tex
\begin{landscape}
\begin{longtable}{lccc}
\caption{\label{tab:K233_MCMC} List of parameters used in the analysis of K2-233. The respective priors are provided together with the posteriors. Only the PARSEC stellar evolution tracks were used. The posterior values represent the median and 68.3\% credible interval. Derived values that might be useful for follow-up work are also reported.}\\
\hline
Parameter & Prior & Posterior \\
 &  &  \\
\hline
\endfirsthead
\multicolumn{4}{l}{{\bfseries \tablename\ \thetable{} -- continued from previous page}} \\
\hline
Parameter & Prior & Posterior\\
 &  &  \\
\hline
\endhead
\multicolumn{3}{l}{{Continued on next page}} \\ 
\hline
\endfoot
\hline
\multicolumn{3}{l}{Notes:}\\
\multicolumn{3}{l}{$\bullet$ $\mathcal{N}(\mu,\sigma^{2})$: Normal distribution with mean $\mu$ and width $\sigma^{2}$}\\
\multicolumn{3}{l}{$\bullet$ $\mathcal{U}(a,b)$: Uniform distribution between $a$ and $b$}\\
\multicolumn{3}{l}{$\bullet$ $\mathcal{S}(a,b)$: Sine distribution between $a$ and $b$, corresponding to $\mathcal{U}(\sin{a},\sin{b})$}\\
%\multicolumn{4}{l}{$\bullet$ $\mathcal{N}_{\mathcal{U}}(\mu,\sigma^{2},a,b)$: Normal distribution with mean $\mu$ and width $\sigma^{2}$ multiplied with a uniform distribution between $a$ and $b$}\\
%\multicolumn{3}{l}{$\bullet$ $\mathcal{S}(a,b)$: Sine distribution between $a$ and $b$}\\
\multicolumn{3}{l}{$\bullet$ $\mathcal{T}(\mu,\sigma^{2},a,b)$: Truncated normal distribution with mean $\mu$ and width $\sigma^{2}$, between $a$ and $b$}\\
\endlastfoot
\\
\multicolumn{3}{l}{\it Stellar Parameters}\\
\\
Initial mass M$_{init}$\ [\Msun] & $\mathcal{N}(0.80,0.05)$ & $0.7929\pm0.0072$  \\
Stellar age $\tau$\ [Gyr] & $\mathcal{U}(0.1,1.0)$  & $0.6\pm0.3$  \\
Iron abundance \met\ [dex] &  $\mathcal{N}(-0.082,0.028)$ & $-0.09\pm0.03$  \\
Distance to Earth $D$ [pc] &  $\mathcal{N}(67.39,0.3)$ & $67.4\pm0.3$  \\
Interstellar extinction $E(B-V)$ [mag] &  $\mathcal{U}(0.0,1.0)$ & $0.031^{_{+0.021}}_{^{-0.018}}$  \\
Systemic radial velocity $\gamma$ [\kms] & $\mathcal{U}(-15.0,-5.0)$ & $-9.6483^{_{+0.0055}}_{^{-0.0056}}$  \\
Effective temperature \teff\ [K] & (derived) & $5032.6^{_{+40.9}}_{^{-34.5}}$  \\
Surface gravity \logg\ [cgs] & (derived) & $4.640\pm0.004$  \\
Linear limb-darkening coefficient $u_{a}$ & (derived) & $0.551^{_{+0.0102}}_{^{-0.0114}}$ \\
Quadratic limb-darkening coefficient $u_{b}$ & (derived) & $0.1653^{_{+0.008}}_{^{-0.0073}}$  \\
Stellar density $\rho_{\star}/\rho_{\astrosun}$ & (derived) & $2.250\pm0.035$  \\
Stellar mass M$_{\star}$\ [\Msun] & (derived) & $0.793\pm0.007$  \\
Stellar radius R$_{\star}$\ [\Rsun] & (derived) & $0.706\pm0.005$  \\

&& \\
\hline
\\
\multicolumn{3}{l}{\it Planet b Parameters}\\
\\
Orbital Period $P_{b}$ [d] &  $\mathcal{N}(2.46746,0.001)$ & $2.46750\pm0.00004$  \\
Transit epoch $T_{0,b}$ [BJD - 2450000] &  $\mathcal{N}(7991.691,0.01)$ & $7991.69112^{_{+0.00083}}_{^{-0.00084}}$  \\
Radial velocity semi-amplitude $K_{b}$ [\ms] & $\mathcal{U}(0,100)$ & $1.86^{_{+1.65}}_{^{-1.22}}$  \\
Orbital inclination $i_{b}$ [$^{\circ}$] & $\mathcal{S}(70.0,90.0)$ & $89.3^{_{+0.5}}_{^{-0.7}}$ \\
Planet-to-star radius ratio $k_{b}$ & $\mathcal{U}(0.0,1.0)$ & $0.01743^{_{+0.00019}}_{^{-0.00016}}$  \\
Orbital eccentricity $e_{b}$ & $\mathcal{T}(0.0,0.083,0.0,1.0)$ & $0.079^{_{+0.048}}_{^{-0.029}}$  \\
Argument of periastron $\omega_{b}$ [\degr] & $\mathcal{U}(0.0,360.0)$ & $248^{_{+53}}_{^{-35}}$  \\
System scale $a_{b}/R_{\star}$ & (derived) & $10.1\pm0.1$ \\
Impact parameter $b_{b}$ & (derived) & $0.128^{_{+0.131}}_{^{-0.089}}$  \\
Transit duration T$_{14,b}$ [h] & (derived) & $1.99\pm0.03$  \\
Semi-major axis $a_{b}$ [AU] & (derived) & $0.03308^{_{+0.00010}}_{^{-0.00009}}$  \\
Planet mass M$_{b}$ [\Mearth] & (derived) & $3.35^{_{+2.97}}_{^{-2.20}}$  \\
Planet radius R$_{b}$ [\Rearth] & (derived) & $1.343^{_{+0.018}}_{^{-0.016}}$  \\
Planet bulk density $\rho_{b}$ [\gcm3] & (derived) & $7.6^{_{+6.7}}_{^{-5.0}}$  \\
&& \\
\hline
\\
\multicolumn{3}{l}{\it Planet c Parameters}\\
\\
Orbital Period $P_{c}$ [d] &  $\mathcal{N}(7.06142,0.01)$ & $7.06005^{_{+0.00024}}_{^{-0.00027}}$  \\
Transit epoch $T_{0,c}$ [BJD - 2450000] &  $\mathcal{N}(7586.9746,0.05)$ & $7586.87653^{_{+0.01769}}_{^{-0.01567}}$  \\
Radial velocity semi-amplitude $K_{c}$ [\ms] & $\mathcal{U}(0,100)$ & $1.99^{_{+1.26}}_{^{-1.14}}$  \\
Orbital inclination $i_{c}$ [$^{\circ}$] & $\mathcal{S}(70.0,90.0)$ & $89.6^{_{+0.3}}_{^{-0.4}}$ \\
Planet-to-star radius ratio $k_{c}$ & $\mathcal{U}(0.0,1.0)$ & $0.01663^{_{+0.00026}}_{^{-0.00024}}$  \\
Orbital eccentricity $e_{c}$ & $\mathcal{T}(0.0,0.083,0.0,1.0)$ & $0.070^{_{+0.045}}_{^{-0.033}}$  \\
Argument of periastron $\omega_{c}$ [\degr] & $\mathcal{U}(0.0,360.0)$ & $261^{_{+52}}_{^{-44}}$  \\
System scale $a_{c}/R_{\star}$ & (derived) & $20.3\pm0.1$  \\
Impact parameter $b_{c}$ & (derived) & $0.133^{_{+0.141}}_{^{-0.093}}$  \\
Transit duration T$_{14,c}$ [h] & (derived) & $2.80^{_{+0.06}}_{^{-0.05}}$  \\
Semi-major axis $a_{c}$ [AU] & (derived) & $0.06666^{_{+0.00020}}_{^{-0.00018}}$  \\
Planet mass M$_{c}$ [\Mearth] & (derived) & $5.09^{_{+3.23}}_{^{-2.91}}$  \\
Planet radius R$_{c}$ [\Rearth] & (derived) & $1.281^{_{+0.022}}_{^{-0.021}}$ \\
Planet bulk density $\rho_{c}$ [\gcm3] & (derived) & $13.3^{_{+8.5}}_{^{-7.6}}$  \\
&& \\
\hline
\\
\multicolumn{3}{l}{\it Planet d Parameters}\\
\\
Orbital Period $P_{d}$ [d] &  $\mathcal{N}(24.3662,0.01)$ & $24.36450\pm0.00068$  \\
Transit epoch $T_{0,d}$ [BJD - 2450000] &  $\mathcal{N}(8005.5801,0.01)$ & $8005.58237^{_{+0.00092}}_{^{-0.00088}}$ \\
Radial velocity semi-amplitude $K_{d}$ [\ms] & $\mathcal{U}(0,100)$ & $2.13^{_{+1.36}}_{^{-1.20}}$  \\
Orbital inclination $i_{d}$ [$^{\circ}$] & $\mathcal{S}(70.0,90.0)$ & $89.6^{_{+0.2}}_{^{-0.1}}$ \\
Planet-to-star radius ratio $k_{d}$ & $\mathcal{U}(0.0,1.0)$ & $0.03062^{_{+0.00048}}_{^{-0.00040}}$  \\
Orbital eccentricity $e_{d}$ & $\mathcal{T}(0.0,0.083,0.0,1.0)$ & $0.062^{_{+0.043}}_{^{-0.042}}$  \\
Argument of periastron $\omega_{d}$ [\degr] & $\mathcal{U}(0.0,360.0)$ & $139^{_{+100}}_{^{-82}}$ \\
System scale $a_{d}/R_{\star}$ & (derived) & $46.4\pm0.2$ \\
Impact parameter $b_{d}$ & (derived) & $0.317^{_{+0.105}}_{^{-0.197}}$ \\
Transit duration T$_{14,d}$ [h] & (derived) & $3.84\pm0.03$  \\
Semi-major axis $a_{d}$ [AU] & (derived) & $0.15224^{_{+0.00046}}_{^{-0.00042}}$  \\
Planet mass M$_{d}$ [\Mearth] & (derived) & $8.25^{_{+5.25}}_{^{-4.66}}$  \\
Planet radius R$_{d}$ [\Rearth] & (derived) & $2.358^{_{+0.043}}_{^{-0.036}}$ \\
Planet bulk density $\rho_{d}$ [\gcm3] & (derived) & $3.4^{_{+2.2}}_{^{-1.9}}$ \\
&& \\
\hline
&& \\
\multicolumn{3}{l}{\it Instrument-related Parameters}\\
&& \\
\hline
&& \\
HARPS jitter $\sigma_{j,~\rm RV}$ [\ms] & $\mathcal{U}(0,100)$ & $0.84\pm0.53$ \\
\textit{K2} contamination [\%] & $\mathcal{T}(0.0,0.005,0.0,1.0)$ & $0.003^{_{+0.004}}_{^{-0.002}}$  \\
\textit{K2} jitter $\sigma_{j,~\textit{K2}}$ [ppm] & $\mathcal{U}(0.0, 10^5)$ & $49.6\pm0.7$ \\
\textit{K2} out-of-transit flux & $\mathcal{U}(0.99,1.01)$ & $0.9999994^{_{+0.0000010}}_{^{-0.0000009}}$  \\
SED jitter [mag]  & $\mathcal{U}(0.0,0.1)$ & $0.036^{_{+0.017}}_{^{-0.012}}$  \\
&& \\
\multicolumn{3}{l}{\it Gaussian Process Hyperparameters}\\
&& \\
A  [\ms] & $\mathcal{U}(0.0,1000)$ & $19.8\pm3.1$ \\
$\lambda_1$ [d] & $\mathcal{U}(15.0,1000)$ & $16.39\pm1.30$  \\
$\lambda_2$ & $\mathcal{U}(0.1, 4)$ & $0.574\pm0.058$ \\
$P_{rot}$ & $\mathcal{N}(10.0,0.5)$ & $9.638\pm0.083$  \\
&& \\
\end{longtable}
\end{landscape}

%% file: Lillo-Box_K2-32_K2-233_PlanetarySystems.bbl
\begin{thebibliography}{80}
\expandafter\ifx\csname natexlab\endcsname\relax\def\natexlab#1{#1}\fi

\bibitem[{{Adibekyan} {et~al.}(2015){Adibekyan}, {Figueira}, {Santos}, {Sousa},
  {Faria}, {Delgado-Mena}, {Oshagh}, {Tsantaki}, {Hakobyan}, {Gonz{\'a}lez
  Hern{\'a}ndez}, {Su{\'a}rez-Andr{\'e}s}, \& {Israelian}}]{adibekyan15}
{Adibekyan}, V., {Figueira}, P., {Santos}, N.~C., {et~al.} 2015, \aap, 583, A94

\bibitem[{{Adibekyan} {et~al.}(2012){Adibekyan}, {Sousa}, {Santos}, {Delgado
  Mena}, {Gonz{\'a}lez Hern{\'a}ndez}, {Israelian}, {Mayor}, \&
  {Khachatryan}}]{adibekyan12}
{Adibekyan}, V.~Z., {Sousa}, S.~G., {Santos}, N.~C., {et~al.} 2012, \aap, 545,
  A32

\bibitem[{{Akeson} {et~al.}(2013){Akeson}, {Chen}, {Ciardi}, {Crane}, {Good},
  {Harbut}, {Jackson}, {Kane}, {Laity}, {Leifer}, {Lynn}, {McElroy}, {Papin},
  {Plavchan}, {Ram{\'{\i}}rez}, {Rey}, {von Braun}, {Wittman}, {Abajian},
  {Ali}, {Beichman}, {Beekley}, {Berriman}, {Berukoff}, {Bryden}, {Chan},
  {Groom}, {Lau}, {Payne}, {Regelson}, {Saucedo}, {Schmitz}, {Stauffer},
  {Wyatt}, \& {Zhang}}]{akeson13}
{Akeson}, R.~L., {Chen}, X., {Ciardi}, D., {et~al.} 2013, \pasp, 125, 989

\bibitem[{{Allard} {et~al.}(2012){Allard}, {Homeier}, \& {Freytag}}]{allard12}
{Allard}, F., {Homeier}, D., \& {Freytag}, B. 2012, Philosophical Transactions
  of the Royal Society of London Series A, 370, 2765

\bibitem[{{Ambikasaran} {et~al.}(2014){Ambikasaran}, {Foreman-Mackey},
  {Greengard}, {Hogg}, \& {O'Neil}}]{george}
{Ambikasaran}, S., {Foreman-Mackey}, D., {Greengard}, L., {Hogg}, D.~W., \&
  {O'Neil}, M. 2014

\bibitem[{{Armstrong} {et~al.}(2019){Armstrong}, {Meru}, {Bayliss}, {Kennedy},
  \& {Veras}}]{armstrong19}
{Armstrong}, D.~J., {Meru}, F., {Bayliss}, D., {Kennedy}, G.~M., \& {Veras}, D.
  2019, \apjl, 880, L1

\bibitem[{{Baranne} {et~al.}(1996){Baranne}, {Queloz}, {Mayor}, {Adrianzyk},
  {Knispel}, {Kohler}, {Lacroix}, {Meunier}, {Rimbaud}, \& {Vin}}]{baranne96}
{Baranne}, A., {Queloz}, D., {Mayor}, M., {et~al.} 1996, \aaps, 119, 373

\bibitem[{{Barros} {et~al.}(2016){Barros}, {Demangeon}, \&
  {Deleuil}}]{barros16}
{Barros}, S.~C.~C., {Demangeon}, O., \& {Deleuil}, M. 2016, \aap, 594, A100

\bibitem[{{Barros} {et~al.}(2017){Barros}, {Gosselin}, {Lillo-Box}, {Bayliss},
  {Delgado Mena}, {Brugger}, {Santerne}, {Armstrong}, {Adibekyan}, {Armstrong},
  {Barrado}, {Bento}, {Boisse}, {Bonomo}, {Bouchy}, {Brown}, {Cochran},
  {Collier Cameron}, {Deleuil}, {Demangeon}, {D{\'\i}az}, {Doyle}, {Dumusque},
  {Ehrenreich}, {Espinoza}, {Faedi}, {Faria}, {Figueira}, {Foxell},
  {H{\'e}brard}, {Hojjatpanah}, {Jackman}, {Lendl}, {Ligi}, {Lovis}, {Melo},
  {Mousis}, {Neal}, {Osborn}, {Pollacco}, {Santos}, {Sefako}, {Shporer},
  {Sousa}, {Triaud}, {Udry}, {Vigan}, \& {Wyttenbach}}]{barros17}
{Barros}, S.~C.~C., {Gosselin}, H., {Lillo-Box}, J., {et~al.} 2017, \aap, 608,
  A25

\bibitem[{{Batalha} {et~al.}(2019){Batalha}, {Lewis}, {Fortney}, {Batalha},
  {Kempton}, {Lewis}, \& {Line}}]{batalha19}
{Batalha}, N.~E., {Lewis}, T., {Fortney}, J.~J., {et~al.} 2019, \apjl, 885, L25

\bibitem[{{Becker} {et~al.}(2015){Becker}, {Vanderburg}, {Adams}, {Rappaport},
  \& {Schwengeler}}]{becker15}
{Becker}, J.~C., {Vanderburg}, A., {Adams}, F.~C., {Rappaport}, S.~A., \&
  {Schwengeler}, H.~M. 2015, \apjl, 812, L18

\bibitem[{{Bertran de Lis} {et~al.}(2015){Bertran de Lis}, {Delgado Mena},
  {Adibekyan}, {Santos}, \& {Sousa}}]{bertran15}
{Bertran de Lis}, S., {Delgado Mena}, E., {Adibekyan}, V.~Z., {Santos}, N.~C.,
  \& {Sousa}, S.~G. 2015, \aap, 576, A89

\bibitem[{{Borucki} {et~al.}(2010){Borucki}, {Koch}, {Basri}, {Batalha},
  {Brown}, {Caldwell}, {Caldwell}, {Christensen-Dalsgaard}, {Cochran},
  {DeVore}, {Dunham}, {Dupree}, {Gautier}, {Geary}, {Gilliland}, {Gould},
  {Howell}, {Jenkins}, {Kondo}, {Latham}, {Marcy}, {Meibom}, {Kjeldsen},
  {Lissauer}, {Monet}, {Morrison}, {Sasselov}, {Tarter}, {Boss}, {Brownlee},
  {Owen}, {Buzasi}, {Charbonneau}, {Doyle}, {Fortney}, {Ford}, {Holman},
  {Seager}, {Steffen}, {Welsh}, {Rowe}, {Anderson}, {Buchhave}, {Ciardi},
  {Walkowicz}, {Sherry}, {Horch}, {Isaacson}, {Everett}, {Fischer}, {Torres},
  {Johnson}, {Endl}, {MacQueen}, {Bryson}, {Dotson}, {Haas}, {Kolodziejczak},
  {Van Cleve}, {Chandrasekaran}, {Twicken}, {Quintana}, {Clarke}, {Allen},
  {Li}, {Wu}, {Tenenbaum}, {Verner}, {Bruhweiler}, {Barnes}, \&
  {Prsa}}]{borucki10}
{Borucki}, W.~J., {Koch}, D., {Basri}, G., {et~al.} 2010, Science, 327, 977

\bibitem[{{Bressan} {et~al.}(2012){Bressan}, {Marigo}, {Girardi}, {Salasnich},
  {Dal Cero}, {Rubele}, \& {Nanni}}]{bressan12}
{Bressan}, A., {Marigo}, P., {Girardi}, L., {et~al.} 2012, \mnras, 427, 127

\bibitem[{Brooks {et~al.}(2003)Brooks, Giudici, \& Philippe}]{brooks03}
Brooks, S., Giudici, P., \& Philippe, A. 2003, Journal of Computational and
  Graphical Statistics, 12

\bibitem[{{Brugger} {et~al.}(2017){Brugger}, {Mousis}, {Deleuil}, \&
  {Deschamps}}]{brugger17}
{Brugger}, B., {Mousis}, O., {Deleuil}, M., \& {Deschamps}, F. 2017, \apj, 850,
  93

\bibitem[{{Buchhave} {et~al.}(2012){Buchhave}, {Latham}, {Johansen},
  {Bizzarro}, {Torres}, {Rowe}, {Batalha}, {Borucki}, {Brugamyer}, {Caldwell},
  {Bryson}, {Ciardi}, {Cochran}, {Endl}, {Esquerdo}, {Ford}, {Geary},
  {Gilliland}, {Hansen}, {Isaacson}, {Laird}, {Lucas}, {Marcy}, {Morse},
  {Robertson}, {Shporer}, {Stefanik}, {Still}, \& {Quinn}}]{buchhave12}
{Buchhave}, L.~A., {Latham}, D.~W., {Johansen}, A., {et~al.} 2012, \nat, 486,
  375

\bibitem[{{Claret} \& {Bloemen}(2011)}]{claret11}
{Claret}, A. \& {Bloemen}, S. 2011, \aap, 529, A75

\bibitem[{{Cresswell} \& {Nelson}(2008)}]{cresswell08}
{Cresswell}, P. \& {Nelson}, R.~P. 2008, \aap, 482, 677

\bibitem[{{Cutri} \& {et al.}(2014)}]{cutri14}
{Cutri}, R.~M. \& {et al.} 2014, VizieR Online Data Catalog, II/328

\bibitem[{{Dai} {et~al.}(2016){Dai}, {Winn}, {Albrecht}, {Arriagada},
  {Bieryla}, {Butler}, {Crane}, {Hirano}, {Johnson}, {Kiilerich}, {Latham},
  {Narita}, {Nowak}, {Palle}, {Ribas}, {Rogers}, {Sanchis-Ojeda}, {Shectman},
  {Teske}, {Thompson}, {Van Eylen}, {Vanderburg}, {Wittenmyer}, \&
  {Yu}}]{dai16}
{Dai}, F., {Winn}, J.~N., {Albrecht}, S., {et~al.} 2016, \apj, 823, 115

\bibitem[{{David} {et~al.}(2018){David}, {Crossfield}, {Benneke}, {Petigura},
  {Gonzales}, {Schlieder}, {Yu}, {Isaacson}, {Howard}, {Ciardi}, {Mamajek},
  {Hillenbrand}, {Cody}, {Riedel}, {Schwengeler}, {Tanner}, \&
  {Ende}}]{david18}
{David}, T.~J., {Crossfield}, I. J.~M., {Benneke}, B., {et~al.} 2018, \aj, 155,
  222

\bibitem[{{Delisle}(2017)}]{delisle17}
{Delisle}, J.~B. 2017, \aap, 605, A96

\bibitem[{{D{\'{\i}}az} {et~al.}(2014){D{\'{\i}}az}, {Almenara}, {Santerne},
  {Moutou}, {Lethuillier}, \& {Deleuil}}]{diaz14b}
{D{\'{\i}}az}, R.~F., {Almenara}, J.~M., {Santerne}, A., {et~al.} 2014, \mnras,
  441, 983

\bibitem[{{Dorn} {et~al.}(2015){Dorn}, {Khan}, {Heng}, {Connolly}, {Alibert},
  {Benz}, \& {Tackley}}]{dorn15}
{Dorn}, C., {Khan}, A., {Heng}, K., {et~al.} 2015, \aap, 577, A83

\bibitem[{{Dorn} {et~al.}(2017){Dorn}, {Venturini}, {Khan}, {Heng}, {Alibert},
  {Helled}, {Rivoldini}, \& {Benz}}]{Dorn17}
{Dorn}, C., {Venturini}, J., {Khan}, A., {et~al.} 2017, \aap, 597, A37

\bibitem[{{Dotter} {et~al.}(2008){Dotter}, {Chaboyer}, {Jevremovi{\'c}},
  {Kostov}, {Baron}, \& {Ferguson}}]{dotter08}
{Dotter}, A., {Chaboyer}, B., {Jevremovi{\'c}}, D., {et~al.} 2008, \apjs, 178,
  89

\bibitem[{{Fulton} {et~al.}(2017){Fulton}, {Petigura}, {Howard}, {Isaacson},
  {Marcy}, {Cargile}, {Hebb}, {Weiss}, {Johnson}, {Morton}, {Sinukoff},
  {Crossfield}, \& {Hirsch}}]{fulton17}
{Fulton}, B.~J., {Petigura}, E.~A., {Howard}, A.~W., {et~al.} 2017, \aj, 154,
  109

\bibitem[{{Gaia Collaboration} {et~al.}(2018){Gaia Collaboration}, {Brown},
  {Vallenari}, {Prusti}, {de Bruijne}, {Babusiaux}, {Bailer-Jones}, {Biermann},
  {Evans}, {Eyer}, {Jansen}, {Jordi}, {Klioner}, {Lammers}, {Lindegren},
  {Luri}, {Mignard}, {Panem}, {Pourbaix}, {Randich}, {Sartoretti}, {Siddiqui},
  {Soubiran}, {van Leeuwen}, {Walton}, {Arenou}, {Bastian}, {Cropper},
  {Drimmel}, {Katz}, {Lattanzi}, {Bakker}, {Cacciari}, {Casta{\~n}eda},
  {Chaoul}, {Cheek}, {De Angeli}, {Fabricius}, {Guerra}, {Holl}, {Masana},
  {Messineo}, {Mowlavi}, {Nienartowicz}, {Panuzzo}, {Portell}, {Riello},
  {Seabroke}, {Tanga}, {Th{\'e}venin}, {Gracia-Abril}, {Comoretto},
  {Garcia-Reinaldos}, {Teyssier}, {Altmann}, {Andrae}, {Audard},
  {Bellas-Velidis}, {Benson}, {Berthier}, {Blomme}, {Burgess}, {Busso},
  {Carry}, {Cellino}, {Clementini}, {Clotet}, {Creevey}, {Davidson}, {De
  Ridder}, {Delchambre}, {Dell'Oro}, {Ducourant},
  {Fern{\'a}ndez-Hern{\'a}ndez}, {Fouesneau}, {Fr{\'e}mat}, {Galluccio},
  {Garc{\'\i}a-Torres}, {Gonz{\'a}lez-N{\'u}{\~n}ez}, {Gonz{\'a}lez-Vidal},
  {Gosset}, {Guy}, {Halbwachs}, {Hambly}, {Harrison}, {Hern{\'a}ndez},
  {Hestroffer}, {Hodgkin}, {Hutton}, {Jasniewicz}, {Jean-Antoine-Piccolo},
  {Jordan}, {Korn}, {Krone-Martins}, {Lanzafame}, {Lebzelter}, {L{\"o}ffler},
  {Manteiga}, {Marrese}, {Mart{\'\i}n-Fleitas}, {Moitinho}, {Mora}, {Muinonen},
  {Osinde}, {Pancino}, {Pauwels}, {Petit}, {Recio-Blanco}, {Richards},
  {Rimoldini}, {Robin}, {Sarro}, {Siopis}, {Smith}, {Sozzetti}, {S{\"u}veges},
  {Torra}, {van Reeven}, {Abbas}, {Abreu Aramburu}, {Accart}, {Aerts},
  {Altavilla}, {{\'A}lvarez}, {Alvarez}, {Alves}, {Anderson}, {Andrei},
  {Anglada Varela}, {Antiche}, {Antoja}, {Arcay}, {Astraatmadja}, {Bach},
  {Baker}, {Balaguer-N{\'u}{\~n}ez}, {Balm}, {Barache}, {Barata}, {Barbato},
  {Barblan}, {Barklem}, {Barrado}, {Barros}, {Barstow}, {Bartholom{\'e}
  Mu{\~n}oz}, {Bassilana}, {Becciani}, {Bellazzini}, {Berihuete}, {Bertone},
  {Bianchi}, {Bienaym{\'e}}, {Blanco-Cuaresma}, {Boch}, {Boeche}, {Bombrun},
  {Borrachero}, {Bossini}, {Bouquillon}, {Bourda}, {Bragaglia}, {Bramante},
  {Breddels}, {Bressan}, {Brouillet}, {Br{\"u}semeister}, {Brugaletta},
  {Bucciarelli}, {Burlacu}, {Busonero}, {Butkevich}, {Buzzi}, {Caffau},
  {Cancelliere}, {Cannizzaro}, {Cantat-Gaudin}, {Carballo}, {Carlucci},
  {Carrasco}, {Casamiquela}, {Castellani}, {Castro-Ginard}, {Charlot},
  {Chemin}, {Chiavassa}, {Cocozza}, {Costigan}, {Cowell}, {Crifo}, {Crosta},
  {Crowley}, {Cuypers}, {Dafonte}, {Damerdji}, {Dapergolas}, {David}, {David},
  {de Laverny}, {De Luise}, {De March}, {de Martino}, {de Souza}, {de Torres},
  {Debosscher}, {del Pozo}, {Delbo}, {Delgado}, {Delgado}, {Di Matteo},
  {Diakite}, {Diener}, {Distefano}, {Dolding}, {Drazinos}, {Dur{\'a}n},
  {Edvardsson}, {Enke}, {Eriksson}, {Esquej}, {Eynard Bontemps}, {Fabre},
  {Fabrizio}, {Faigler}, {Falc{\~a}o}, {Farr{\`a}s Casas}, {Federici},
  {Fedorets}, {Fernique}, {Figueras}, {Filippi}, {Findeisen}, {Fonti},
  {Fraile}, {Fraser}, {Fr{\'e}zouls}, {Gai}, {Galleti}, {Garabato},
  {Garc{\'\i}a-Sedano}, {Garofalo}, {Garralda}, {Gavel}, {Gavras}, {Gerssen},
  {Geyer}, {Giacobbe}, {Gilmore}, {Girona}, {Giuffrida}, {Glass}, {Gomes},
  {Granvik}, {Gueguen}, {Guerrier}, {Guiraud}, {Guti{\'e}rrez-S{\'a}nchez},
  {Haigron}, {Hatzidimitriou}, {Hauser}, {Haywood}, {Heiter}, {Helmi}, {Heu},
  {Hilger}, {Hobbs}, {Hofmann}, {Holland}, {Huckle}, {Hypki}, {Icardi},
  {Jan{\ss}en}, {Jevardat de Fombelle}, {Jonker}, {Juh{\'a}sz}, {Julbe},
  {Karampelas}, {Kewley}, {Klar}, {Kochoska}, {Kohley}, {Kolenberg},
  {Kontizas}, {Kontizas}, {Koposov}, {Kordopatis}, {Kostrzewa-Rutkowska},
  {Koubsky}, {Lambert}, {Lanza}, {Lasne}, {Lavigne}, {Le Fustec}, {Le
  Poncin-Lafitte}, {Lebreton}, {Leccia}, {Leclerc}, {Lecoeur-Taibi},
  {Lenhardt}, {Leroux}, {Liao}, {Licata}, {Lindstr{\o}m}, {Lister}, {Livanou},
  {Lobel}, {L{\'o}pez}, {Managau}, {Mann}, {Mantelet}, {Marchal}, {Marchant},
  {Marconi}, {Marinoni}, {Marschalk{\'o}}, {Marshall}, {Martino}, {Marton},
  {Mary}, {Massari}, {Matijevi{\v{c}}}, {Mazeh}, {McMillan}, {Messina},
  {Michalik}, {Millar}, {Molina}, {Molinaro}, {Moln{\'a}r}, {Montegriffo},
  {Mor}, {Morbidelli}, {Morel}, {Morris}, {Mulone}, {Muraveva}, {Musella},
  {Nelemans}, {Nicastro}, {Noval}, {O'Mullane}, {Ord{\'e}novic},
  {Ord{\'o}{\~n}ez-Blanco}, {Osborne}, {Pagani}, {Pagano}, {Pailler},
  {Palacin}, {Palaversa}, {Panahi}, {Pawlak}, {Piersimoni}, {Pineau}, {Plachy},
  {Plum}, {Poggio}, {Poujoulet}, {Pr{\v{s}}a}, {Pulone}, {Racero}, {Ragaini},
  {Rambaux}, {Ramos-Lerate}, {Regibo}, {Reyl{\'e}}, {Riclet}, {Ripepi}, {Riva},
  {Rivard}, {Rixon}, {Roegiers}, {Roelens}, {Romero-G{\'o}mez}, {Rowell},
  {Royer}, {Ruiz-Dern}, {Sadowski}, {Sagrist{\`a} Sell{\'e}s}, {Sahlmann},
  {Salgado}, {Salguero}, {Sanna}, {Santana-Ros}, {Sarasso}, {Savietto},
  {Schultheis}, {Sciacca}, {Segol}, {Segovia}, {S{\'e}gransan}, {Shih},
  {Siltala}, {Silva}, {Smart}, {Smith}, {Solano}, {Solitro}, {Sordo}, {Soria
  Nieto}, {Souchay}, {Spagna}, {Spoto}, {Stampa}, {Steele},
  {Steidelm{\"u}ller}, {Stephenson}, {Stoev}, {Suess}, {Surdej}, {Szabados},
  {Szegedi-Elek}, {Tapiador}, {Taris}, {Tauran}, {Taylor}, {Teixeira},
  {Terrett}, {Teyssand ier}, {Thuillot}, {Titarenko}, {Torra Clotet}, {Turon},
  {Ulla}, {Utrilla}, {Uzzi}, {Vaillant}, {Valentini}, {Valette}, {van Elteren},
  {Van Hemelryck}, {van Leeuwen}, {Vaschetto}, {Vecchiato}, {Veljanoski},
  {Viala}, {Vicente}, {Vogt}, {von Essen}, {Voss}, {Votruba}, {Voutsinas},
  {Walmsley}, {Weiler}, {Wertz}, {Wevers}, {Wyrzykowski}, {Yoldas},
  {{\v{Z}}erjal}, {Ziaeepour}, {Zorec}, {Zschocke}, {Zucker}, {Zurbach}, \&
  {Zwitter}}]{gaia18}
{Gaia Collaboration}, {Brown}, A.~G.~A., {Vallenari}, A., {et~al.} 2018, \aap,
  616, A1

\bibitem[{{Hara} {et~al.}(2019){Hara}, {Bou{\'e}}, {Laskar}, {Delisle}, \&
  {Unger}}]{hara19}
{Hara}, N.~C., {Bou{\'e}}, G., {Laskar}, J., {Delisle}, J.~B., \& {Unger}, N.
  2019, \mnras, 489, 738

\bibitem[{{H{\'e}brard} {et~al.}(2014){H{\'e}brard}, {Delfosse}, {Morin},
  {Boisse}, {Moutou}, \& {H{\'e}brard}}]{hebrard14}
{H{\'e}brard}, E.~M., {Delfosse}, X., {Morin}, J., {et~al.} 2014, in SF2A-2014:
  Proceedings of the Annual meeting of the French Society of Astronomy and
  Astrophysics, ed. J.~{Ballet}, F.~{Martins}, F.~{Bournaud}, R.~{Monier}, \&
  C.~{Reyl{\'e}}, 241--244

\bibitem[{{Heller} {et~al.}(2019){Heller}, {Rodenbeck}, \& {Hippke}}]{heller19}
{Heller}, R., {Rodenbeck}, K., \& {Hippke}, M. 2019, \aap, 625, A31

\bibitem[{{Hellier} {et~al.}(2012){Hellier}, {Anderson}, {Collier Cameron},
  {Doyle}, {Fumel}, {Gillon}, {Jehin}, {Lendl}, {Maxted}, {Pepe}, {Pollacco},
  {Queloz}, {S{\'e}gransan}, {Smalley}, {Smith}, {Southworth}, {Triaud},
  {Udry}, \& {West}}]{hellier12}
{Hellier}, C., {Anderson}, D.~R., {Collier Cameron}, A., {et~al.} 2012, \mnras,
  426, 739

\bibitem[{{Henden} {et~al.}(2015){Henden}, {Levine}, {Terrell}, \&
  {Welch}}]{henden15}
{Henden}, A.~A., {Levine}, S., {Terrell}, D., \& {Welch}, D.~L. 2015, in
  American Astronomical Society Meeting Abstracts, Vol. 225, American
  Astronomical Society Meeting Abstracts \#225, 336.16

\bibitem[{{Howell} {et~al.}(2014){Howell}, {Sobeck}, {Haas}, {Still},
  {Barclay}, {Mullally}, {Troeltzsch}, {Aigrain}, {Bryson}, {Caldwell},
  {Chaplin}, {Cochran}, {Huber}, {Marcy}, {Miglio}, {Najita}, {Smith},
  {Twicken}, \& {Fortney}}]{howell14}
{Howell}, S.~B., {Sobeck}, C., {Haas}, M., {et~al.} 2014, \pasp, 126, 398

\bibitem[{{Kipping}(2010)}]{kipping10}
{Kipping}, D.~M. 2010, \mnras, 408, 1758

\bibitem[{{Kipping}(2014)}]{kipping14}
{Kipping}, D.~M. 2014, \mnras, 440, 2164

\bibitem[{{Koll} {et~al.}(2019){Koll}, {Malik}, {Mansfield}, {Kempton}, {Kite},
  {Abbot}, \& {Bean}}]{koll19}
{Koll}, D. D.~B., {Malik}, M., {Mansfield}, M., {et~al.} 2019, \apj, 886, 140

\bibitem[{{Kurucz}(1993)}]{Kurucz-1993}
{Kurucz}, R. 1993, ATLAS9 Stellar Atmosphere Programs and 2 km/s grid.~Kurucz
  CD-ROM No.~13.~ Cambridge, Mass.: Smithsonian Astrophysical Observatory,
  1993., 13

\bibitem[{{Lam} {et~al.}(2018){Lam}, {Santerne}, {Sousa}, {Vigan}, {Armstrong},
  {Barros}, {Brugger}, {Adibekyan}, {Almenara}, {Delgado Mena}, {Dumusque},
  {Barrado}, {Bayliss}, {Bonomo}, {Bouchy}, {Brown}, {Ciardi}, {Deleuil},
  {Demangeon}, {Faedi}, {Foxell}, {Jackman}, {King}, {Kirk}, {Ligi},
  {Lillo-Box}, {Lopez}, {Lovis}, {Louden}, {Nielsen}, {McCormac}, {Mousis},
  {Osborn}, {Pollacco}, {Santos}, {Udry}, \& {Wheatley}}]{lam18}
{Lam}, K.~W.~F., {Santerne}, A., {Sousa}, S.~G., {et~al.} 2018, \aap, 620, A77

\bibitem[{{Leleu} {et~al.}(2019{\natexlab{a}}){Leleu}, {Coleman}, \&
  {Ataiee}}]{leleu19b}
{Leleu}, A., {Coleman}, G., \& {Ataiee}, S. 2019{\natexlab{a}}, arXiv e-prints,
  arXiv:1901.07640

\bibitem[{{Leleu} {et~al.}(2019{\natexlab{b}}){Leleu}, {Lillo-Box}, {Sestovic},
  {Robutel}, {Correia}, {Hara}, {Angerhausen}, {Grimm}, \&
  {Schneider}}]{leleu19}
{Leleu}, A., {Lillo-Box}, J., {Sestovic}, M., {et~al.} 2019{\natexlab{b}},
  arXiv e-prints

\bibitem[{{Lillo-Box} {et~al.}(2018){Lillo-Box}, {Barrado}, {Figueira},
  {Leleu}, {Santos}, {Correia}, {Robutel}, \& {Faria}}]{lillo-box18a}
{Lillo-Box}, J., {Barrado}, D., {Figueira}, P., {et~al.} 2018, \aap, 609, A96

\bibitem[{{Lissauer} {et~al.}(2014){Lissauer}, {Marcy}, {Bryson}, {Rowe},
  {Jontof-Hutter}, {Agol}, {Borucki}, {Carter}, {Ford}, {Gilliland}, {Kolbl},
  {Star}, {Steffen}, \& {Torres}}]{lissauer14}
{Lissauer}, J.~J., {Marcy}, G.~W., {Bryson}, S.~T., {et~al.} 2014, \apj, 784,
  44

\bibitem[{{Lopez} {et~al.}(2019){Lopez}, {Barros}, {Santerne}, {Deleuil},
  {Adibekyan}, {Almenara}, {Armstrong}, {Brugger}, {Barrado}, {Bayliss},
  {Boisse}, {Bonomo}, {Bouchy}, {Brown}, {Carli}, {Demangeon}, {Dumusque},
  {D{\'\i}az}, {Faria}, {Figueira}, {Foxell}, {Giles}, {H{\'e}brard},
  {Hojjatpanah}, {Kirk}, {Lillo-Box}, {Lovis}, {Mousis}, {da N{\'o}brega},
  {Nielsen}, {Neal}, {Osborn}, {Pepe}, {Pollacco}, {Santos}, {Sousa}, {Udry},
  {Vigan}, \& {Wheatley}}]{lopez19}
{Lopez}, T.~A., {Barros}, S.~C.~C., {Santerne}, A., {et~al.} 2019, arXiv
  e-prints, arXiv:1909.13527

\bibitem[{{Luger} {et~al.}(2016){Luger}, {Agol}, {Kruse}, {Barnes}, {Becker},
  {Foreman-Mackey}, \& {Deming}}]{luger16}
{Luger}, R., {Agol}, E., {Kruse}, E., {et~al.} 2016, \aj, 152, 100

\bibitem[{{Luger} {et~al.}(2018){Luger}, {Kruse}, {Foreman-Mackey}, {Agol}, \&
  {Saunders}}]{luger18}
{Luger}, R., {Kruse}, E., {Foreman-Mackey}, D., {Agol}, E., \& {Saunders}, N.
  2018, \aj, 156, 99

\bibitem[{{Malik} {et~al.}(2019){Malik}, {Kempton}, {Koll}, {Mansfield},
  {Bean}, \& {Kite}}]{malik19}
{Malik}, M., {Kempton}, E. M.~R., {Koll}, D. D.~B., {et~al.} 2019, \apj, 886,
  142

\bibitem[{{Mansfield} {et~al.}(2019){Mansfield}, {Kite}, {Hu}, {Koll}, {Malik},
  {Bean}, \& {Kempton}}]{mansfield19}
{Mansfield}, M., {Kite}, E.~S., {Hu}, R., {et~al.} 2019, \apj, 886, 141

\bibitem[{{Mayo} {et~al.}(2018){Mayo}, {Vanderburg}, {Latham}, {Bieryla},
  {Morton}, {Buchhave}, {Dressing}, {Beichman}, {Berlind}, {Calkins}, {Ciardi},
  {Crossfield}, {Esquerdo}, {Everett}, {Gonzales}, {Hirsch}, {Horch}, {Howard},
  {Howell}, {Livingston}, {Patel}, {Petigura}, {Schlieder}, {Scott}, {Schumer},
  {Sinukoff}, {Teske}, \& {Winters}}]{mayo18}
{Mayo}, A.~W., {Vanderburg}, A., {Latham}, D.~W., {et~al.} 2018, \aj, 155, 136

\bibitem[{{Mayor} {et~al.}(2003){Mayor}, {Pepe}, {Queloz}, {Bouchy},
  {Rupprecht}, {Lo Curto}, {Avila}, {Benz}, {Bertaux}, {Bonfils}, {Dall},
  {Dekker}, {Delabre}, {Eckert}, {Fleury}, {Gilliotte}, {Gojak}, {Guzman},
  {Kohler}, {Lizon}, {Longinotti}, {Lovis}, {Megevand}, {Pasquini}, {Reyes},
  {Sivan}, {Sosnowska}, {Soto}, {Udry}, {van Kesteren}, {Weber}, \&
  {Weilenmann}}]{mayor03}
{Mayor}, M., {Pepe}, F., {Queloz}, D., {et~al.} 2003, The Messenger, 114, 20

\bibitem[{{Montet} {et~al.}(2015){Montet}, {Morton}, {Foreman-Mackey},
  {Johnson}, {Hogg}, {Bowler}, {Latham}, {Bieryla}, \& {Mann}}]{montet15}
{Montet}, B.~T., {Morton}, T.~D., {Foreman-Mackey}, D., {et~al.} 2015, \apj,
  809, 25

\bibitem[{{Mortier} {et~al.}(2014){Mortier}, {Sousa}, {Adibekyan},
  {Brand{\~a}o}, \& {Santos}}]{mortier14}
{Mortier}, A., {Sousa}, S.~G., {Adibekyan}, V.~Z., {Brand{\~a}o}, I.~M., \&
  {Santos}, N.~C. 2014, \aap, 572, A95

\bibitem[{{Morton}(2012)}]{morton12}
{Morton}, T.~D. 2012, \apj, 761, 6

\bibitem[{{Mousis} {et~al.}(2020){Mousis}, {Deleuil}, {Aguichine}, {Marcq},
  {Naar}, {Acu{\~n}a Aguirre}, {Brugger}, \& {Goncalves}}]{mousis20}
{Mousis}, O., {Deleuil}, M., {Aguichine}, A., {et~al.} 2020, arXiv e-prints,
  arXiv:2002.05243

\bibitem[{{Munari} {et~al.}(2014){Munari}, {Henden}, {Frigo}, {Zwitter},
  {Bienaym{\'e}}, {Bland-Hawthorn}, {Boeche}, {Freeman}, {Gibson}, {Gilmore},
  {Grebel}, {Helmi}, {Kordopatis}, {Levine}, {Navarro}, {Parker}, {Reid},
  {Seabroke}, {Siebert}, {Siviero}, {Smith}, {Steinmetz}, {Templeton},
  {Terrell}, {Welch}, {Williams}, \& {Wyse}}]{munari14}
{Munari}, U., {Henden}, A., {Frigo}, A., {et~al.} 2014, \aj, 148, 81

\bibitem[{{Obermeier} {et~al.}(2016){Obermeier}, {Henning}, {Schlieder},
  {Crossfield}, {Petigura}, {Howard}, {Sinukoff}, {Isaacson}, {Ciardi},
  {David}, {Hillenbrand}, {Beichman}, {Howell}, {Horch}, {Everett}, {Hirsch},
  {Teske}, {Christiansen}, {L{\'e}pine}, {Aller}, {Liu}, {Saglia},
  {Livingston}, \& {Kluge}}]{obermeier16}
{Obermeier}, C., {Henning}, T., {Schlieder}, J.~E., {et~al.} 2016, \aj, 152,
  223

\bibitem[{{Pepe} {et~al.}(2010){Pepe}, {Cristiani}, {Rebolo Lopez}, {Santos},
  {Amorim}, {Avila}, {Benz}, {Bonifacio}, {Cabral}, {Carvas}, {Cirami},
  {Coelho}, {Comari}, {Coretti}, {De Caprio}, {Dekker}, {Delabre}, {Di
  Marcantonio}, {D'Odorico}, {Fleury}, {Garc{\'\i}a}, {Herreros Linares},
  {Hughes}, {Iwert}, {Lima}, {Lizon}, {Lo Curto}, {Lovis}, {Manescau},
  {Martins}, {M{\'e}gevand}, {Moitinho}, {Molaro}, {Monteiro}, {Monteiro},
  {Pasquini}, {Mordasini}, {Queloz}, {Rasilla}, {Rebord{\~a}o}, {Santana
  Tschudi}, {Santin}, {Sosnowska}, {Span{\`o}}, {Tenegi}, {Udry}, {Vanzella},
  {Viel}, {Zapatero Osorio}, \& {Zerbi}}]{espresso}
{Pepe}, F.~A., {Cristiani}, S., {Rebolo Lopez}, R., {et~al.} 2010, Society of
  Photo-Optical Instrumentation Engineers (SPIE) Conference Series, Vol. 7735,
  {ESPRESSO: the Echelle spectrograph for rocky exoplanets and stable
  spectroscopic observations}, 77350F

\bibitem[{{Pepper} {et~al.}(2017){Pepper}, {Gillen}, {Parviainen},
  {Hillenbrand}, {Cody}, {Aigrain}, {Stauffer}, {Vrba}, {David}, {Lillo-Box},
  {Stassun}, {Conroy}, {Pope}, \& {Barrado}}]{pepper17}
{Pepper}, J., {Gillen}, E., {Parviainen}, H., {et~al.} 2017, \aj, 153, 177

\bibitem[{{Petigura} {et~al.}(2017){Petigura}, {Sinukoff}, {Lopez},
  {Crossfield}, {Howard}, {Brewer}, {Fulton}, {Isaacson}, {Ciardi}, {Howell},
  {Everett}, {Horch}, {Hirsch}, {Weiss}, \& {Schlieder}}]{petigura17}
{Petigura}, E.~A., {Sinukoff}, E., {Lopez}, E.~D., {et~al.} 2017, \aj, 153, 142

\bibitem[{{Santerne} {et~al.}(2018){Santerne}, {Brugger}, {Armstrong},
  {Adibekyan}, {Lillo-Box}, {Gosselin}, {Aguichine}, {Almenara}, {Barrado},
  {Barros}, {Bayliss}, {Boisse}, {Bonomo}, {Bouchy}, {Brown}, {Deleuil},
  {Delgado Mena}, {Demangeon}, {D{\'\i}az}, {Doyle}, {Dumusque}, {Faedi},
  {Faria}, {Figueira}, {Foxell}, {Giles}, {H{\'e}brard}, {Hojjatpanah},
  {Hobson}, {Jackman}, {King}, {Kirk}, {Lam}, {Ligi}, {Lovis}, {Louden},
  {McCormac}, {Mousis}, {Neal}, {Osborn}, {Pepe}, {Pollacco}, {Santos},
  {Sousa}, {Udry}, \& {Vigan}}]{santerne18}
{Santerne}, A., {Brugger}, B., {Armstrong}, D.~J., {et~al.} 2018, Nature
  Astronomy, 2, 393

\bibitem[{{Santerne} {et~al.}(2019){Santerne}, {Malavolta}, {Kosiarek}, {Dai},
  {Dressing}, {Dumusque}, {Hara}, {Lopez}, {Mortier}, {Vanderburg},
  {Adibekyan}, {Armstrong}, {Barrado}, {Barros}, {Bayliss}, {Berardo},
  {Boisse}, {Bonomo}, {Bouchy}, {Brown}, {Buchhave}, {Butler}, {Collier
  Cameron}, {Cosentino}, {Crane}, {Crossfield}, {Damasso}, {Deleuil}, {Delgado
  Mena}, {Demangeon}, {D{\'\i}az}, {Donati}, {Figueira}, {Fulton}, {Ghedina},
  {Harutyunyan}, {H{\'e}brard}, {Hirsch}, {Hojjatpanah}, {Howard}, {Isaacson},
  {Latham}, {Lillo-Box}, {L{\'o}pez-Morales}, {Lovis}, {Martinez Fiorenzano},
  {Molinari}, {Mousis}, {Moutou}, {Nava}, {Nielsen}, {Osborn}, {Petigura},
  {Phillips}, {Pollacco}, {Poretti}, {Rice}, {Santos}, {S{\'e}gransan},
  {Shectman}, {Sinukoff}, {Sousa}, {Sozzetti}, {Teske}, {Udry}, {Vigan},
  {Wang}, {Watson}, {Weiss}, {Wheatley}, \& {Winn}}]{santerne19}
{Santerne}, A., {Malavolta}, L., {Kosiarek}, M.~R., {et~al.} 2019, arXiv
  e-prints, arXiv:1911.07355

\bibitem[{{Santos} {et~al.}(2017){Santos}, {Adibekyan}, {Dorn}, {Mordasini},
  {Noack}, {Barros}, {Delgado-Mena}, {Demangeon}, {Faria}, {Israelian}, \&
  {Sousa}}]{santos17}
{Santos}, N.~C., {Adibekyan}, V., {Dorn}, C., {et~al.} 2017, \aap, 608, A94

\bibitem[{{Santos} {et~al.}(2015){Santos}, {Adibekyan}, {Mordasini}, {Benz},
  {Delgado-Mena}, {Dorn}, {Buchhave}, {Figueira}, {Mortier}, {Pepe},
  {Santerne}, {Sousa}, \& {Udry}}]{santos15}
{Santos}, N.~C., {Adibekyan}, V., {Mordasini}, C., {et~al.} 2015, \aap, 580,
  L13

\bibitem[{{Santos} {et~al.}(2013){Santos}, {Sousa}, {Mortier}, {Neves},
  {Adibekyan}, {Tsantaki}, {Delgado Mena}, {Bonfils}, {Israelian}, {Mayor}, \&
  {Udry}}]{santos13}
{Santos}, N.~C., {Sousa}, S.~G., {Mortier}, A., {et~al.} 2013, \aap, 556, A150

\bibitem[{{Schmitt} {et~al.}(2014){Schmitt}, {Agol}, {Deck}, {Rogers}, {Gazak},
  {Fischer}, {Wang}, {Holman}, {Jek}, {Margossian}, {Omohundro}, {Winarski},
  {Brewer}, {Giguere}, {Lintott}, {Lynn}, {Parrish}, {Schawinski}, {Schwamb},
  {Simpson}, \& {Smith}}]{schmitt14}
{Schmitt}, J.~R., {Agol}, E., {Deck}, K.~M., {et~al.} 2014, \apj, 795, 167

\bibitem[{{Sch{\"o}nrich} {et~al.}(2019){Sch{\"o}nrich}, {McMillan}, \&
  {Eyer}}]{schonrich19}
{Sch{\"o}nrich}, R., {McMillan}, P., \& {Eyer}, L. 2019, \mnras, 487, 3568

\bibitem[{{Seager} \& {Sasselov}(2000)}]{seager00}
{Seager}, S. \& {Sasselov}, D.~D. 2000, \apj, 537, 916

\bibitem[{{Sinukoff} {et~al.}(2016){Sinukoff}, {Howard}, {Petigura},
  {Schlieder}, {Crossfield}, {Ciardi}, {Fulton}, {Isaacson}, {Aller},
  {Baranec}, {Beichman}, {Hansen}, {Knutson}, {Law}, {Liu}, {Riddle}, \&
  {Dressing}}]{sinukoff16}
{Sinukoff}, E., {Howard}, A.~W., {Petigura}, E.~A., {et~al.} 2016, \apj, 827,
  78

\bibitem[{{Sneden}(1973)}]{Sneden-1973}
{Sneden}, C.~A. 1973, PhD thesis, THE UNIVERSITY OF TEXAS AT AUSTIN.

\bibitem[{{Sousa}(2014)}]{sousa14}
{Sousa}, S.~G. 2014, ArXiv e-prints 1407.5817S

\bibitem[{{Sousa} {et~al.}(2018){Sousa}, {Adibekyan}, {Delgado-Mena}, {Santos},
  {Andreasen}, {Ferreira}, {Tsantaki}, {Barros}, {Demangeon}, {Israelian},
  {Faria}, {Figueira}, {Mortier}, {Brandao}, {Montalto}, {Rojas-Ayala}, \&
  {Santerne}}]{Sousa-2018}
{Sousa}, S.~G., {Adibekyan}, V., {Delgado-Mena}, E., {et~al.} 2018, ArXiv
  e-prints

\bibitem[{{Sousa} {et~al.}(2015){Sousa}, {Santos}, {Adibekyan}, {Delgado-Mena},
  \& {Israelian}}]{sousa15}
{Sousa}, S.~G., {Santos}, N.~C., {Adibekyan}, V., {Delgado-Mena}, E., \&
  {Israelian}, G. 2015, \aap, 577, A67

\bibitem[{{Sousa} {et~al.}(2007){Sousa}, {Santos}, {Israelian}, {Mayor}, \&
  {Monteiro}}]{Sousa-2007}
{Sousa}, S.~G., {Santos}, N.~C., {Israelian}, G., {Mayor}, M., \& {Monteiro},
  M.~J.~P.~F.~G. 2007, A\&A, 469, 783

\bibitem[{{Southworth}(2008)}]{southworth08}
{Southworth}, J. 2008, \mnras, 386, 1644

\bibitem[{{Su{\'a}rez-Andr{\'e}s} {et~al.}(2017){Su{\'a}rez-Andr{\'e}s},
  {Israelian}, {Gonz{\'a}lez Hern{\'a}ndez}, {Adibekyan}, {Delgado Mena},
  {Santos}, \& {Sousa}}]{suarez-andres17}
{Su{\'a}rez-Andr{\'e}s}, L., {Israelian}, G., {Gonz{\'a}lez Hern{\'a}ndez},
  J.~I., {et~al.} 2017, \aap, 599, A96

\bibitem[{{Tsantaki} {et~al.}(2013){Tsantaki}, {Sousa}, {Adibekyan}, {Santos},
  {Mortier}, \& {Israelian}}]{Tsantaki-2013}
{Tsantaki}, M., {Sousa}, S.~G., {Adibekyan}, V.~Z., {et~al.} 2013, \aap, 555,
  A150

\bibitem[{{Van Eylen} {et~al.}(2019){Van Eylen}, {Albrecht}, {Huang},
  {MacDonald}, {Dawson}, {Cai}, {Foreman-Mackey}, {Lundkvist}, {Silva Aguirre},
  {Snellen}, \& {Winn}}]{vaneylen19}
{Van Eylen}, V., {Albrecht}, S., {Huang}, X., {et~al.} 2019, \aj, 157, 61

\bibitem[{{Weiss} {et~al.}(2018){Weiss}, {Marcy}, {Petigura}, {Fulton},
  {Howard}, {Winn}, {Isaacson}, {Morton}, {Hirsch}, {Sinukoff}, {Cumming},
  {Hebb}, \& {Cargile}}]{weiss18}
{Weiss}, L.~M., {Marcy}, G.~W., {Petigura}, E.~A., {et~al.} 2018, \aj, 155, 48

\bibitem[{{Zeng} {et~al.}(2019){Zeng}, {Jacobsen}, {Sasselov}, {Petaev},
  {Vanderburg}, {Lopez-Morales}, {Perez-Mercader}, {Mattsson}, {Li}, {Heising},
  {Bonomo}, {Damasso}, {Berger}, {Cao}, {Levi}, \& {Wordsworth}}]{zeng19}
{Zeng}, L., {Jacobsen}, S.~B., {Sasselov}, D.~D., {et~al.} 2019, Proceedings of
  the National Academy of Science, 116, 9723

\end{thebibliography}
